\documentclass[a4paper,12pt]{report} 
\usepackage[utf8]{inputenc} 
\pdfoutput=1
\usepackage{amsmath,amsfonts,amssymb,amsthm,mathbbol} 
\usepackage[T1]{fontenc} 
\usepackage[utf8]{inputenc} 
\usepackage{mathtools}
\usepackage{breqn}
 
\usepackage[hidelinks]{hyperref} 
\hypersetup{ colorlinks=false}
\usepackage{graphicx}
\usepackage{cancel}
\usepackage{cleveref}
\usepackage{tikz}
\usepackage{caption}
\usepackage{enumitem} 
\usepackage{ragged2e}
\usepackage{authblk}
\usepackage{labelschanged}
\usepackage[legalpaper, portrait]{geometry}
 \usepackage[tracking=true]{microtype}
\DeclareMicrotypeSet*[tracking]{my}
 { encoding = *,
 size = {-small,Large-},
 font = */*/*/sc/* }
\SetTracking[ no ligatures = f ]{ encoding = *, shape = sc}{ 50 }
\SetTracking{ encoding = *, size = -small }{ 20 }
\SetTracking{ encoding = *, size = Large- }{ -20 }
\newcommand{\stirlingii}{\genfrac{\{}{\}}{0pt}{}}
 
\makeatletter
\newsavebox\myboxA
\newsavebox\myboxB
\newlength\mylenA

\newcommand*\xoverline[2][0.75]{%
    \sbox{\myboxA}{$\m@th#2$}%
    \setbox\myboxB\null
    \ht\myboxB=\ht\myboxA%
    \dp\myboxB=\dp\myboxA%
    \wd\myboxB=#1\wd\myboxA
    \sbox\myboxB{$\m@th\overline{\copy\myboxB}$}
    \setlength\mylenA{\the\wd\myboxA}
    \addtolength\mylenA{-\the\wd\myboxB}%
    \ifdim\wd\myboxB<\wd\myboxA%
       \rlap{\hskip 0.5\mylenA\usebox\myboxB}{\usebox\myboxA}%
    \else
        \hskip -0.5\mylenA\rlap{\usebox\myboxA}{\hskip 0.5\mylenA\usebox\myboxB}%
    \fi}
\makeatother
\title{Double layers in Weyl geometry and some physical implications}
\author[1]{Victor Berezin \thanks{\href{mailto:berezin@inr.ac.ru}{berezin@inr.ac.ru}}}
\author[2]{Kushan Dharmadarshi \thanks{\href{mailto:kushan.d@iitgn.ac.in}{kushan.d@iitgn.ac.in}}}
\affil[1]{Institute for Nuclear Research of the Russian Academy of Sciences, 60th
October Anniversary Prospect, 7a, 117312, Moscow, Russia}

\begin{document} 
\maketitle

\chapter*{Conventions} This report for the 
most part utilises the 
\href{https://en.wikipedia.org/wiki/Sign_convention#Relativity}{\textbf{
Landau-Lifshitz Spacelike Convention}}\label{conv::link}, as popularised by 
Misner, Wheeler and Thorne \cite{MTW1973}; 
\begin{align*} {{R}^{\mu}}_{\nu\rho\sigma} = 
{{\partial}_{\rho}}{\Gamma^{\mu}_{\nu\sigma}} - 
{{\partial}_{\sigma}}{\Gamma^{\mu}_{\nu\rho}} + 
{\Gamma^{\mu}_{\lambda\rho}}{\Gamma^{\lambda}_{\nu\sigma}} - 
{\Gamma^{\mu}_{\lambda\sigma}}{\Gamma^{\lambda}_{\nu\rho}} 
\end{align*} The 
Ricci tensor is defined as: 
\begin{align*} R_{\mu\nu} \equiv 
{R^{\lambda}}_{\mu\lambda\nu} 
\end{align*} The commutator with its relevant 
conventions: 
\begin{align*} [\nabla_{\mu},\nabla_{\nu}]{V^{\lambda}} &= 
{{R^{\lambda}}_{\rho\mu\nu}}{V^{\rho}}\\ 
[\nabla_{\mu},\nabla_{\nu}]{h^{\alpha\beta}} &= 
{h^{\beta\lambda}}{R^{\alpha}}_{\lambda\mu\nu} + {h^{\alpha\lambda}}{R^{\beta}}_{\lambda\mu\nu} 
\end{align*} 
 All Latin letters $i,j,k,l,\cdots$ represent three dimensional coordinates, whilst their Greek counterparts $\alpha ,\beta ,\gamma ,\delta ,\mu ,\nu ,\cdots$ represent 4-dimensional ones.\newline
\textbf{Derivative Conventions}\\
We have three distinct types of derivatives which are used in this paper;
\begin{enumerate}[label= \roman*.]
 \item ${{R^\mu}_{\nu\lambda\sigma}}_{,\alpha}$ is the partial derivative with respect to the $\alpha$ index
 \item ${{R^\mu}_{\nu\lambda\sigma}}_{;\alpha}$ is the covariant derivative, with respect to the four dimensional Christoffel's symbols.
 \item ${{R^\mu}_{\nu\lambda\sigma}}_{|\alpha}$ is the three dimensional derivative, with respect to the four dimensional Christoffel's symbols.
\end{enumerate}
The signature we will use for our calculations: $(+,-,-,-)$

\tableofcontents 

\chapter{A Brief Introduction} 
\label{chap:intro} 
\normalsize 
So far in my dealings with the premise of my work, I have spent some time studying Weyls 
original attempt (in 1918) at the unification of Maxwells electrodynamics with Einsteins Gravity theory (the General Theory of Relativity as later christened by the Royal Society). \\Previous attempts had been made to achieve the same, specifically Mies attempt \cite{Corry2018}, from which Weyl drew a lot of his inspiration, albeit his attempt, although ultimately not acceptable because of physical repercussions (the second clock effect and other misgivings as pointed out by Einstein, Pauli and other contemporaries \cite{Vizgin1994}), was mathematically ingenious, indeed. \\A very brief introduction to the same can be found below (and in detail in \cite{ROMERO2019180}), I shall introduce the basic formalism here.\\
Weyl geometry is perhaps one of the simplest generalisations of Riemannian geometry, the only modification being the fact that the covariant derivative isn't metric compatible (i.e the covariant derivative of the metric tensor isn't zero), but instead given by:
\begin{align}
{\nabla}_{\alpha} {g_{\beta\lambda}} = {\sigma_{\alpha}}{g_{\beta\lambda}}  \label{eq:ch1}
\end{align}
Weyl believed that a true unified theory of the present fundamental forces at the time (Gravity and Electromagnetic/Lorentz force(s)), could never be achieved in the framework of Riemannian geometry, which was at the base of GR. He based his effort in the realm of his own "Pure infinitesimal geometry" \cite{ScholzInfGeometry}, which took a more general view point as compared to Riemannian geometry, and didn't allow for the comparison of directions between any two points on a manifold as was afforded by Riemannian geometry. Instead, his version involved a comparison in vector quantities only at points belonging to an infinitesimal neighbourhood. This led to length calibration at every event on a manifold and further to the inclusion of "Length Curvature", to add to the normal sense of curvature due to a disparity in autoparallely transported tangent vectors on a given curve on the manifold. 
Where ${\sigma_{\alpha}}$ denotes the components of a one-form field $\sigma$ in a local coordinate basis. This weakening of the Riemannian compatibility condition is entirely equivalent to requiring that the length of a vector field may change when parallel-transported along a curve in the manifold. We refer to the triple $(M,g,\sigma)$  consisting of a differentiable manifold M endowed with both a metric $g$ and a 1-form field $\sigma$ as a Weyl gauge (or, frame). Now one important discovery made by Weyl was the following. Suppose we perform the conformal transformation:
\begin{align}
\overline{g} = e^f g \label{eq:ch2}
\end{align}
where $f$ is an arbitrary scalar function defined on $M$. Then, the Weyl compatibility condition \eqref{eq:ch1} still holds provided that we let the Weyl field $\sigma$ transform as:
\begin{align}
\overline{\sigma} = \sigma + df \label{eq:ch3}
\end{align}
In other words, the Weyl compatibility condition does not change when we go from one gauge $(M,g,\sigma)$ to another gauge $(M,\overline{g},\overline{\sigma})$ by simultaneously transformations in $g$ and $\sigma$.
If we assume that the Weyl connection $\nabla$ is symmetric, a straightforward algebra shows that one can express the components of the affine connection in an arbitrary vector basis completely in terms of the components of $g$ and $\sigma$:
\begin{align}
{\Gamma^{\alpha}_{\beta\lambda}} = \stirlingii{\alpha}{\beta \hspace{2mm} \lambda} - \frac{1}{2}{g^{\alpha\mu}}\left\lbrace{{g_{\mu\beta}}{A_{\lambda}} + {g_{\mu\lambda}}{A_{\beta}} -{g_{\beta\lambda}}{A_{\mu}}}\right\rbrace \label{eq:ch4}
\end{align}
In \eqref{eq:4}, the $\stirlingii{\alpha}{\beta \hspace{2mm} \lambda}$ represents the Christoffel symbols, which describe the Riemannian connection of GR.\\
This clearly the changes the playing field. Fiddling around with his theory, Weyl posited the existence of a new 2-form $F_{\mu\nu}$, which he related to the antisymmetric difference between the partial derivatives of the 1-form $\sigma_{\alpha}$, this is what he correlated to the vector 4-potential of the Electromagnetic field and further is strikingly similar to the gauging of the Electromagnetic potential. \\
This is the basic premise of Weyls Unified theory. In the years to come since his initial suggestion of it, he received admiration for the mathematical beauty of the theory, albeit also severe critique over its failings as a cohesive theory of nature.\\
In the first three chapters \ref{chap:pre}, \ref{chap:pig} and \ref{chap:first}, I attempt to describe the developments that took place in the early twentieth century in the realm of Unified Field Theories, focusing especially on Weyls Unification of GR and classical Maxwellian Electrodynamics. The following chapters elucidate upon the same, deriving from a wide range of references \cite{Goenner2004}, \cite{Scholz2019tif}, \cite{ScholzInfGeometry}, \cite{ROMERO2019180}, \cite{Vizgin1994} and \cite{AdlerSchiffer} and the reader is directed to them to explore the subject in greater detail. A basic understanding of General Relativity Theory is assumed.
In the final two chapters, \ref{chap:DLT} and \ref{chap:conclude}, I dive into investigating the properties of the Weyl Quadratic gravity that sets it apart from GR as well as novel effects that the coveted Double Layer exhibits under a few different considerations, especially in terms on the surface energy tensor $S_{ni}$ which is theoretically observed in Double Layers in Weyl gravity.
\chapter{The Impetus}
\label{chap:impetus}
We study Weyl Quadratic gravity (quadratic for reasons that become apparent in chapter \ref{chap:first}), in order to be able to describe physical phenomena related to the Double Layer. The Double Layer is a unique manifestation of the discontinuities in the curvature tensor (as is shown in chapter \ref{chap:DLT}). It is claimed that they exhibit special surface energy tensor components which can manifest in many interesting ways, unlike those observed in GR, from which they are completely absent.\\
We also wish to find out a relationship of sorts, if it should exist, between the gauging one-form (more on that in the coming chapters) and any entity local to the Double Layer itself. This will help us better understand the true nature of the gauging one-form, rather than it just being a device for failed unification, a purpose which it inadvertently served in Weyls attempt at unifying Gravity and Electrodynamics (see chapter \ref{chap:first}).
\chapter{Geometrical Preliminaries} 
\label{chap:pre}
Before we begin our foray into a most exciting time in physics, let us first establish, briefly a geometrical foreground which shall aid us in understanding the theory which is to be discussed later.\\
We begin with;\newline
\textbf{\underline{The Metric}}\newline
We are all privy to the notion of the line element:
\begin{align}
ds = \sqrt{{g_{ij}} d x^i d x^j}
\label{eq:1}
\end{align}
We are used to considering a symmetric $g$ (metric) tensor field, with $\frac{D(D+1)}{2}$ components, where $D$ is the dimension of the manifold under consideration.\\
this metric is said to be "Riemannian", if its Eigenvalues are positive (negative) definite and Lorentzian if its signature is $\pm (D-2)$.\\
Although the Riemannian metric taken in GR, is symmetric, here that will not hold and the metric may have a decomposition of the kind:
\begin{align}
{g_{ik}} = {\gamma_{(ik)}} + {\phi_{[ik]}}\label{eq:2}
\end{align}
Where, ${\gamma_{(ik)}}$ and ${\phi_{[ik]}}$ denote the symmetric and asymmetric parts of the metric $g_{ik}$ respectively. The inverse would consequently be:
\begin{align}
{g^{ik}} = {h^{(ik)}} + {f^{[ik]}}\label{eq:3}
\end{align}
These are further related to each other as given:
\begin{align}
{\gamma_{ij}}{\gamma^{ik}} = {\delta^k_j}; \hspace{2mm} {\phi_{ij}}{\phi^{ik}} = {\delta^k_j}; \hspace{2mm} {h_{ij}}{h^{ik}} = {\delta^k_j}; \hspace{2mm} {f_{ij}}{f^{ik}} = {\delta^k_j}; \hspace{2mm}\label{eq:4}
\end{align}
\begin{align}
{h^{(ik)}} = \frac{\gamma}{g} {\gamma^{ik}} + \frac{\phi}{g} {\phi^{im}}{\phi^{kn}} {\gamma_{mn}}\label{eq:5}
\end{align}
\begin{align}
{f^{(ik)}} = \frac{\phi}{g} {\phi^{ik}} + \frac{\gamma}{g} {\gamma^{im}}{\gamma^{kn}} {\phi_{mn}}\label{eq:6}
\end{align}
Here $g, \phi, \gamma$ represent $det(g_{ik})$, $det(\phi_{ik})$ and $det(\gamma_{ik})$ respectively.
\begin{align}
 g = \gamma + \phi + \frac{\gamma}{2} {\gamma^{kl}}{\gamma^{mn}} {\phi_{km}}{\phi_{ln}}\label{eq:7}
\end{align}
A manifold is called "\textit{spacetime}", if $D=4$, the metric is symmetric and Lorentzian (symmetric and with signature $\text{sig}(g) = \pm 2$). Albeit often, this definition is not adhered to, with higher dimensional manifolds equipped with geometrical structures besides those listed here being connoted as "\textit{spacetime}", albeit here, we shall adhere to this definition as prescribed.
\newpage
 
The metric tensor may also be described through D-vector fields forming an orthonormal D-leg (-bein) \cite{bergmann_sabbata_1986} $h^k_{\hat{i}}$ with:
\begin{align}
{g_{lm}} = {h_{i{\hat{j}}}}{h_{m{\hat{k}}}}{\eta^{{\hat{j}}{\hat{k}}}}\label{eq:eq8}
\end{align}
The hatted indices count the number of legs spanning the tangent space at each point ($\hat{j} = 1,2,...,D$); these are transformed using the Minkowski metric. we can further introduce 1-forms (ala Cartan), of the sort:
\begin{align}
{\theta^{\hat{k}}} = {h^{\hat{k}}_l} d x^l\label{eq:9}
\end{align}
And then write the square of our line element as follows:
\begin{align}
d s^2 = {\theta^{\hat{i}}}{\theta^{\hat{k}}} {\eta_{{\hat{i}}{\hat{k}}}}\label{eq:10}
\end{align}
Next, we have;\newline
\textbf{\underline{Affine structure}}\newline
Affine structure, often described with the aid of a metric (but not exclusively so, depending on the geometrical prescription of the theory), is a \textit{linear connection $\mathcal{L}$} with $D^3$ components ${L_{ij}}^k$. It is a
geometrical object but not a tensor field and its components change inhomogeneously under local
coordinate transformations. using this \textit{linear connection}, we may define a \textit{covariant derivative}, as follows:
\begin{align}
{\overset{+}{\nabla}}_k X^i = \frac{\partial X^i}{\partial x^k} + {L_{kj}}^i X^j ; \hspace{2mm} {\overset{+}{\nabla}}_k \omega^i = \frac{\partial \omega^i}{\partial x^k} - {L_{ki}}^j \omega^j \label{eq:11}
\end{align}
\begin{align}
{\overset{-}{\nabla}}_k X^i = \frac{\partial X^i}{\partial x^k} + {L_{jk}}^i X^j ; \hspace{2mm} {\overset{-}{\nabla}}_k \omega^i = \frac{\partial \omega^i}{\partial x^k} - {L_{ik}}^j \omega^j \label{eq:12}
\end{align}
Here the change in the appearance of the \textit{covariant derivative} (with the negative sign on top) in \eqref{eq:12} is to denote a change in the order of the indices of the \textit{connection}. If the connection is symmetric, the difference between these two conventions is naught and we can simplify the covariant derivative  to $\nabla$, instead of the ones used in \eqref{eq:11} and \eqref{eq:12}. If the connection isn't symmetric, we get:
\begin{align}
{\overset{+}{\nabla}}_k X^i - {\overset{-}{\nabla}}_k X^i = 2{L_{[kj]}}^i X^j\label{eq:13}
\end{align}
As the difference.\\
A manifold equipped with only a linear connection is called as \textit{Affine space}. With respect to the Affine group (group of inhomogenous coordinate transformations), the connection transforms like a tensor would.\\
For a vector density $\hat{X}$, the covariant derivative is:
\begin{align}
{\overset{+}{\nabla}}_k X^i = \frac{\partial {\hat{X}}^i}{\partial x^k} + {L_{kj}}^i {\hat{X}}^j - {L_{kr}}^r {\hat{X}}^i; \hspace{2mm} {\overset{-}{\nabla}}_k X^i = \frac{\partial {\hat{X}}^i}{\partial x^k} + {L_{jk}}^i {\hat{X}}^j - {L_{rk}}^r {\hat{X}}^i\label{eq:14}
\end{align}
Transformation mapping autoparallels to autoparallels:
\begin{align}
{L_{ik}}^j \rightarrow {L_{ik}}^j + {{\delta^j}_{(i}}{\omega_{k)}}\label{eq:15}
\end{align}
The equivalence class of autoparallels defined in this way (as done in \eqref{eq:15}), is said to define a \textit{Projective Structure} on the manifold ($M_D$).
This particular set of connections:
\begin{align}
_{(p)}{L_{ij}}^k = {L_{ij}}^k - \frac{2}{D+1} {{\delta^k}_{(i} {L_{j)}}}\label{eq:16}
\end{align}
with $L_{j} = {L_{jl}}^l$ is mapped into itself by \eqref{eq:15}.\\
From connection ${L_{ij}}^k$ further connections can be constructed, by adding an arbitrary tensor field to its symmetrised part. This is how we can arrive at other connections, starting by using a given connection as the baseline.
\begin{align}
{{\overline{L}}_{ij}}^k = {L_{(ij)}}^k + {T_{ij}}^k = {\Gamma_{ij}}^k + {T_{ij}}^k\label{eq:17}
\end{align}
The asymmetric part is given by:
\begin{align}
 {S_{ij}}^k  = {L_{[ij]}}^k = {T_{[ij]}}^k \rightarrow \text{Torsion, tensor field}\label{eq:18}
\end{align}
$S_{j} = {S_{jl}}^l \rightarrow$ Torsion vector, connecting two traces of the affine connection:
\begin{align}
L_{j} = {L_{jl}}^l; \hspace{2mm} {\widetilde{L}}_{j} = {{L}_{lj}}^l; \hspace{2mm} \text{as}\hspace{2mm} S_i = \frac{1}{2} \left( L_j - {\widetilde{L}}_{j}\right)\label{eq:19}
\end{align}
Having been re-acquainted with these two basic structures, we can discuss the different types of geometries based on them;\newline
\textbf{\underline{Affine Geometry}}\newline
Various subcases of affine geometry can occur, depending on whether the connection used is asymmetric or symmetric. In affine geometry, the metric is derived solely with the help of the connection, or rather the tensorial objects derived from it. This is in stark contrast to Riemannian geometry, where the connection is derived from the metric. These tensorial objects are the curvature tensors, (Two of them for the negative and positive symmetry convention as detailed in \eqref{eq:11} and \eqref{eq:12}):
\begin{align}
{\stackrel{+}{K}^i}_{jkl} = \partial_k {L_{lj}}^i - \partial_l {L_{kj}}^i + {L_{km}}^i {L_{lj}}^m - {L_{lm}}^i {L_{kj}}^m \label{eq:20}
\end{align}
\begin{align}
{\stackrel{-}{K}^i}_{jkl} = \partial_k {L_{jl}}^i - \partial_l {L_{jk}}^i + {L_{mk}}^i {L_{jl}}^m - {L_{ml}}^i {L_{jk}}^m \label{eq:21}
\end{align}
In a geometry with a symmetric affine connection, both of them coincide.
\begin{align}
\frac{1}{2} \left({\stackrel{+}{K}^i}_{jkl} - {\stackrel{-}{K}^i}_{jkl} \right) = \delta_{[k} {S_{]lj}}^i  + 2{S_{j[k}}^m {S_{l]m}}^i + {L_{m[k}}^i {S_{l]j}}^m - {L_{j[k}}^m {S_{l]m}}^i\label{eq:22}
\end{align}
The curvature tensor arises by virtue of the covariant derivative not being Abelian and adhering to the Ricci identity;
\begin{align}
{\overset{+}{\nabla}}_{[j} {\overset{+}{\nabla}}_{k]} A^i = \frac{1}{2} {\stackrel{+}{K}^i}_{rjk} A^r - {S_{jk}}^r {\overset{+}{\nabla}}_{r} A^i\label{eq:23}
\end{align}
\begin{align}
{\overset{-}{\nabla}}_{[j} {\overset{-}{\nabla}}_{k]} A^i = \frac{1}{2} {\stackrel{-}{K}^i}_{rjk} A^r + {S_{jk}}^r {\overset{-}{\nabla}}_{r} A^i\label{eq:24}
\end{align}
For vector density, the identity is given by: 
\begin{align}
{\overset{+}{\nabla}}_{[j} {\overset{+}{\nabla}}_{k]} \hat{A}^i = \frac{1}{2} {\stackrel{+}{K}^i}_{rjk} A^r - {S_{jk}}^r {\overset{+}{\nabla}}_{r} \hat{A}^i + \frac{1}{2} {V_{jk}} \hat{A}^i\label{eq:25}
\end{align}
The $V_{jk}$ defined here is called \textit{homothetic curvature} and is given as:
\begin{align}
{V_{kl}} = \partial_k L_l - \partial_l L_k\label{eq:26}
\end{align}
It is also synonymous with the tensorial trace of affine curvature $V_{kl} = {K^i}_{ikl} = {V_{[kl]}}$
The symmetries hence observed by the curvature tensor are:
\begin{align}
{\stackrel{-}{K}^i}_{j[kl]}  &= 0\\ 
                                    \label{eq:27}
{\stackrel{-}{K}^i}_{\lbrace jkl \rbrace} &= 2 {\nabla_{\lbrace j}} {S_{kl\rbrace}}^i = 4 {S_{m\lbrace j}}^i {S_{kl\rbrace }}^m
\end{align}
Where the $\lbrace \rbrace$ indicate cyclic permutation.
The Bianchi identity applies as;
\begin{align}
{\stackrel{+}{K^i}}_{j\lbrace lk || m\rbrace} =  2 {k^i}_{r\lbrace kl} {S_{m\rbrace j}}^r\label{eq:29}
\end{align}
Where the pipeline symbol "$||$" denotes a covariant derivative, with respect to the succeeding parameter.
And further identities: 
\begin{align}
{V_{kl}} + 2{K_{[kl]}} = 4 {\nabla_{[k}}{S_{l]}} + 8{S_{kl}}^m {S_m} + 2 {\nabla_{m}} {S_{kl}}^m \label{eq:30}
\end{align}
\begin{align}
{{\xoverline{V}}_{kl}} + 2{{\xoverline{K}}_{[kl]}} = -4 {{\xoverline{\nabla}}_{[k}}{S_{l]}} + 8{S_{kl}}^m {S_m} + 2 {\nabla_{m}} {S_{kl}}^m \label{eq:31}
\end{align}
Symmetric and asymmetric parts, $K_{jk}$
\begin{align}
{K_{[kl]}} = -{\partial_{[k}}{{\widetilde{L}}_{l]}} + {\nabla_m}{S_{kl}}^m + {L_{m}}{S_{kl}}^m + 2 {L_{[l|r}}^{m} {S_{m|k]}}^r\label{eq:32}
\end{align}
\begin{align}
{K_{[kl]}} = {\partial_{[k}}{{\widetilde{L}}_{l]}} - {\partial_m}{L_{(kl)}}^m - {{\widetilde{L}}_{m}}{L_{(kl)}}^m + {L_{(k|m|}}^{n} {L_{l)n}}^m\label{eq:33}
\end{align}
Here the single pipeline notation "$|$" in \eqref{eq:32} and \eqref{eq:33}, is used to exclude $k$ from the symmetrisation bracket.
For a symmetric tensor (i.e ${S_{kl}}^m = 0$), we have:
\begin{align}
{{K}^i}_{\lbrace jkl \rbrace} = 0\label{eq:34}
\end{align}
\begin{align}
{{K^i}}_{j\lbrace lk || m\rbrace} = 0\label{eq:35}
\end{align}
\begin{align}
{V_{kl}} + 2{K_{[kl]}} = 0\label{eq:36}
\end{align}
These expressions are the usual ones which we observe in Riemannian curvature symmetries.
Only one independent trace tensor of the affine curvature tensor exists. While for the asymmetric part of the Ricci tensor;
\begin{align}
{K_{[kl]}} = - {\partial_{[k}} {{\widetilde{L}}_{l]}} \label{eq:37}
\end{align}
holds.\\
The easiest method of defining the fundamental form/tensor, is by setting: $g_{ij} = \alpha K_{(ij)}$ or $g_{ij} = \alpha {\xoverline{K}}_{(ij)}$. Or, alternatively, to derive the metric from the simplest Lagrangian scalar density that could used as such, given by $\text{det}(K_{ij})$.\\
Curvature tensor as calculated from the connection, $\xoverline{L}_{ij} = {\Gamma_{ij}}^k + {T_{ij}}^k$, expressed by curvature tensor of ${\Gamma_{ij}}^k$ and tack on the tensor ${T_{ij}}^k$:
\begin{align}
{{K}^i}_{\lbrace jkl \rbrace} (\xoverline{L}) = {{K}^i}_{\lbrace jkl \rbrace} (\Gamma) + 2 ^{(\Gamma)}{\nabla_{[k}}{T_{l]j}}^{i} - 2 {T_{[k|j|}}^m {T_{l]m}}^i + 2 {S_{kl}}^m {T_{mj}}^i\label{eq:38}
\end{align}
Here the $^{(\Gamma)}{\nabla}$ represents a covariant derivative derived utilising the ${\Gamma_{ij}}^k$ connection.
We now move on to;\newline
\textbf{\underline{Mixed Geometry}}\newline
A manifold carrying both structural elements, i.e. the metric and the connection is referred to as a \textit{metric-affine space}. If the first fundamental is taken to be asymmetric, i.e. to "contain" an asymmetric part: ${g_{[ij]}} = \frac{1}{2} \left({g_{ij}} - {g_{ji}}\right)$, then we call this specialised case, \textit{mixed geometry}, this can be interpreted as Riemannian geometry with additional structures (geometrical objects/entities); such as the 2-form field $\phi (f)$ (symplectic form), Torsion $S$ and non-metricity $Q$, depending on the physical interpretation most well-suited to represent the relationship shared amongst mathematical entities on the manifold and their physically observable counterparts.\\
As is well known from literature, from the symmetric part of the of the first fundamental form, $h_{ij} = g_{(ij)}$, a connection may be constructed, ergo the \textit{Levi-Civita} connection:
\begin{align}
{\stirlingii{k}{i \hspace{2mm} j}} = \frac{1}{2} {\gamma^{kl}} {\left( {\gamma_{li,j}} + {\gamma_{lj,i}} - {\gamma_{ij,l}}\right)}\label{eq:39} 
\end{align}
With the Riemannian curvature tensor formally defined, with ${L_{ij}}^{k} = {\stirlingii{k}{i \hspace{2mm} j}}$ and with the help of the symmetric affine connection, the non-metricity tensor ${Q_{ij}}^k$ can now be given as:
\begin{align}
{Q_{ij}}^k = {g^{kl}} {\nabla_{l}} {g_{ij}}\label{eq:40}
\end{align}
Then, we get the connection for this mixed geometry as:
\begin{align}
{{\Gamma}_{ij}}^k = {\stirlingii{k}{i \hspace{2mm} j}} + {{K_{ij}}^k} + \frac{1}{2} {\left\lbrace {{Q^{k}}_{ij}} +{{Q_{ji}}^k} - {{{Q_{j}}^{k}}_{i}}\right\rbrace}\label{eq:41}
\end{align}
The cotorsion tensor: 
\begin{align}
{K_{ij}}^k = {{S^{k}}_{ji}} + {{S^{k}}_{ij}} - {{S_{ij}}^k} = - {{{{K_{i}}}^k}_j}\label{eq:42}
\end{align}
Inner product of 2 tangent vectors, $A^i$,$B^k$ is not conserved under parallel transport of them, along $X^l$ if the non-metricity tensor, does not vanish:
\begin{align}
{X^k} \overset{+}{\nabla}_{k} ({A^n}{B^m} {g_{nm}}) = {Q_{nml}}{A^n}{B^m}{X^l} \neq 0 \label{eq:43}
\end{align}
The connection for which non-metricity tensor vanishes:
\begin{align}
{\overset{+}{\nabla}_{j}}{g_{ik}} = 0\label{eq:44}
\end{align}
i.e. it is metric compatible.\\
J.H Thomas believed that a combination of $\overset{+}{\nabla}$ and $\overset{-}{\nabla}$, to create a covariant derivative for the metric:
\begin{align}
{g_{ik|l}} = {g_{ik,l}} - {g_{rk}}{\Gamma^{r}_{il}} - {g_{ir}}{\Gamma^{r}_{lk}}\label{eq:45} 
\end{align} 
Extended it for tensors of arbitrary rank $\geq 3$.\\
Einstein used as a constraint on the metric tensor:
\begin{align}
0 = \underset{+-}{g_{ik||l}} = {g_{ik,l}} - {g_{rk}}{\Gamma^{r}_{il}} - {g_{ir}}{\Gamma^{r}_{lk}}\label{eq:46}
\end{align}
In Weyl theory, we have:
\begin{align}
{Q_{ij|k}} = {Q_k}{g_{ij}} \rightarrow \text{Semi-metrical} \label{eq:47}
\end{align}
Inner product multiplies by scalar factor during parallel transport:
\begin{align}
{X^k}{\overset{+}{\nabla}_k} ({A^n}{B^m} {g_{nm}}) = ({Q_l}{X^l}) {A^n}{B^m} {g_{nm}}\label{eq:48}
\end{align}
Light cone is further preserved by parallel transport along $X^l$.\\
If we define: ${{X_{ij}}^k} = {{Q^k}_{ij}} + {{Q_{ji}}^k} - {{{{Q}_{j}}^k}_i}$
Curvature Tensor of a Torsionless affine space given by:
\begin{align}
{{K^{i}}_{jkl}} (\overline{\Gamma}) = {{K^{i}}_{jkl}} \left(\stirlingii{r}{n \hspace{2mm} m}\right) + 2 ^{\left(\stirlingii{i}{j \hspace{2mm} k}\right)}{\nabla_{[k}}{{X_{l]j}}^i} - 2{{X_{[k|j}}^m}{{X_{l]m}}^i}\label{eq:49}
\end{align}
$^{\left(\stirlingii{i}{j \hspace{2mm} k}\right)}{\nabla}$, covariant derivative formed from Christoffel's Symbol.\\
Riemann-Cartan geometry subcase of metric affine geometry, in which metric compatible connection $\rightarrow$ contains Torsion i.e. an asymmetric part ${L_{[ij]}}^k$, Torsion is a tensor field $\rightarrow$ linked to physical observables. For a linear connection with asymmetric part, ${S_{ij}}^k$:
\begin{align}
{S_{ij}}^k = {S_{[i}} {\delta_{j]}}^k \rightarrow \text{Semi-symmetric}\label{eq:50}
\end{align}
Riemannian geometry, further subcase with vanishing torsion of a metric-affine geometry with metric-compatible connection.\\
In this case, connection derived from metric: ${\Gamma_{ij}}^k =  \stirlingii{i}{j \hspace{2mm} k}$
in metric-affine and in mixed geometry, geodesic and autoparallel, curves will have to be distinguished.\\
With a conformal transformation:
\begin{align}
{g_{ik}}  \rightarrow {g_{ik}}^\prime = \lambda {g_{ik}}\label{eq:51}
\end{align}
Smooth function also changes the components of the non-metricity tensor:
\begin{align}
{{Q_{ij}}^k} \rightarrow {{Q_{ij}}^k} + {g_{ij}}{g^{kl}}{\partial_{l}}{\sigma}\label{eq:52}
\end{align}
$\sigma_i = \partial_i \text{log}\lambda$; Riemann curvature tensor, ${R^i_{jkl}} \rightarrow$ also changes.
If ${{R^{\prime i}}_{jkl}} = 0 \rightarrow$ possible through a conformal transformation, in $M_D$; $D>3$, the vanishing of the Weyl curvature tensor: 
\begin{align}
{{C^i}_{jkl}} = {{R^i}_{jkl}} + {\frac{2}{D-2}} \left( {{\delta^{i}}_{[k}} {R_{l]j}} + {g_{j[l}}{{R^{i}}_{k]}}\right)+ {\frac{2R}{(D-1)(D-2)}} {{\delta^{i}}_{[l}} {g_{k]j}}\label{eq:53}
\end{align}
Which is necessary and sufficient condition for $M_D$ to be conformally flat. For $D = 4$, we have:
\begin{align}
{{C^i}_{jkl}} = {{R^i}_{jkl}} +  \left( {{\delta^{i}}_{[k}} {R_{l]j}} + {g_{j[l}}{{R^{i}}_{k]}}\right)+ {\frac{R}{3}} {{\delta^{i}}_{[l}} {g_{k]j}}\label{eq:54}
\end{align}
With the geometrical preliminaries thus established, we can continue to talk about how Weyls theory came into existence and how it was received by his contemporaries.
\chapter{Weyls Pure Infinitesimal Geometry}
\label{chap:pig}
The next step on our voyage through the origins of Weyls Unification, would be to understand the impetus behind the theory per se. We shall discuss the same in this chapter. For a more thorough discussion of the same I would refer the reader to sources such as: \cite{folland1970}\\

Quoting \cite{ScholzInfGeometry} directly:\\
From the perspective of Weyl's view of the continuum around 1918, differential
geometry may be considered as one line of access to the question of how to link
the infinitesimal "halos" of the point to the structure of the whole, the manifold.
From this point of view, the differential geometric structures should be defined such
that only relations in each infinitesimal neighborhood are immediately meaningful.
Relations between quantities in different neighborhoods (of finite distance) ought
to be considered meaningful only by mediation of the whole, more technically
expressed by an integration process over paths joining the two points at the centers
of the neighborhoods. From the point of view of building the continuum from
its smallest parts, so Weyl claimed over and over again in the years following
1918, Riemann's differential geometry did not appear completely convincing. In
Riemannian geometry the relationship between lengths of vectors ~ and T) is well
defined, independent of the points p and q of the manifold to which they are
attached ($\xi \in {T_{p} M}, \hspace{1mm} \eta \in {T_{q} M}, p \neq q$). From Weyl's continuum-based view of
differential geometry such a comparison appeared unmotivated, and he stipulated
instead that a\\
\begin{center}
\justify
{... (a) truly infinitesimal geometry (wahrhafte nahegeometrie) ... should
know a transfer principle for length measurements between infinitely
close points only. [W2, p. 30]
In this formulation Weyl alluded to Levi-Civita's transfer principle of di-
rection in a Riemannian manifold embedded in a sufficiently high-dimensional
Euclidean space, locally given by:} 
\end{center}
\begin{align}
{\xi^{\prime i}} = {\xi^{i}} - {{\Gamma^{i}}_{jk}}{\xi^{j}}{d x^{k}}\label{eq:57}
\end{align}
With the $d x^{i}$ to be interpreted as the coordinate representation of a displacement
vector between two infinitesimally close points so that the direction vector $\xi^{i}$ has been transferred to ${\xi^{\prime i}}$.\\
Weyl immediately recognized that Levi-Civita's concept of parallel displace-
ment wonderfully suited his nascent ideas about how to build differential geometry
strictly on the basis of infinitesimal neighborhoods. He discussed it in this light
during his lecture on general relativity in the summer of 1917 at Zurich, not yet
knowing how to proceed similarly for the measurement and comparison of lengths.
This was the motivation behind Weyl's effort to separate logically the concept
of parallel displacement from metrics and to introduce what he called an affine
connection $\Gamma$ on a (differentiable) manifold as, speaking in later terminology, a
linear torsion-free connection. Guided by the example of affine connections Weyl
also proceeded to build up the metric in a manifold from a "purely infinitesimal
view". The result was his introduction of a generalized Riemannian metric, a
Weylian metric on a differentiable manifold $M$, which is given by:
\begin{enumerate}
 \item A conformal structure on $M$, i.e. a class of (semi-) Riemannian metrics $[g]$ in
local coordinates given by ${g_{ij}} (x)$ or ${{\widetilde{g}}_{ij}} (x) = \lambda (x) {g_{ij}} (x)$, with multiplication
by $\lambda (x) > 0$ (real valued) representing what Weyl considered to be a gauge
transformation of the representative of $[g]$, and
\item A length connection on $M$, i.e. a class of differential 1-forms $\phi$ in local coordi-
nates represented by ${\phi_{i}}{d x_{i}}$, ${{\widetilde{\phi}}_{i}}{d x_{i}} = {\phi_{i}}{d x_{i}} - d{\text{log}}{\lambda}$ (representing the gauge
transformation of the representative of $\phi$).
\end{enumerate}
To put this in more simpler physical terms, in the general theory of relativity, Einstein used Riemannian geometry as a model for physical space. However, the universe is not really a Riemannian manifold, for there is no absolute measure of length, that is, instead of being given a scalar product on the tangent space at each point, we are given a scalar product determined only up to a positive factor at each point. This fact produces no essential change in the geometry provided that a determination of length at one point uniquely induces a determination of length on the whole manifold, i.e., if it makes sense to compare the size of two tangent vectors at two distinct points. Weyl conjectured that this is not the case rather, that an analogy should be drawn with the theory of linear connections, in which it generally makes sense to say that two vectors at two distinct points have the same direction only if there is specified a curve between the two points along which "parallel translation" can take place. Hence in the Weyl theory a determination of length at one point induces only a first-order approximation to a determination of length at surrounding points. Using the ideas established here and in chapter \ref{chap:pre}, we can finally understand Weyls motivations, which we do in the next chapter.

\chapter{Weyls first attempt at unification}
\label{chap:first}
Weyls fundamental idea for generalising Riemannian geometry, as mentioned in the previous chapter was that; unlike  the comparison of vectors at different points of a manifold, a similar comparison between scalars does not elicit the need for a connection.\\
Lengths of vectors at different points can be compared to each other, without a connection, the same does not hold for directions.\\
Weyl thought that a metrical relationship from point to point is only infused into a manifold, if the principle for carrying unit length from one point to its infinitesimal neighbours is given.\\
In Riemannian geometry contrastingly, the line elements can be compared not just infinitesimally but at extended, arbitrary distances.\\
In a "pure infinitesimal geometry", comparison at distance isn't admissible.\\
As mentioned in \ref{chap:intro}, to introduce a pure infinitesimal geometry, Weyl introduced the idea of a 1-dimensional Abelian group of Gauge transformations:
\begin{align}
g \rightarrow \xoverline{g} = \lambda g\label{eq:58}
\end{align}
On top of the diffeomorphism group.\\
This induces a local recalibration of lengths whilst preserving angles ($\delta l = \lambda l$) at each event on the manifold under consideration. Non-metricity tensor is assumed to have a special form $Q_{ijk} = {Q_{k}} {g_{ij}}$.
Then from \eqref{eq:52}, we can obtain:
\begin{align}
{Q_{k}} \rightarrow {Q_{k}} + {\partial_{k}}{\sigma}\label{eq:59}
\end{align}
Which is highly reminiscent of the electromagnetic gauge transformations of the 4-vector potential in Maxwells theory.
If the connection is assumed to be symmetric then we have vanishing torsion and using \eqref{eq:41}, we get:
\begin{align}
{{\Gamma}_{ij}}^k = \stirlingii{k}{i \hspace{2mm} j} + \frac{1}{2} {\left\lbrace {\delta^{k}_{i}}{Q_{j}} +{\delta^{k}_{j}}{Q_{i}} - {g_{ij}}{g^{kl}}{Q_{l}}\right\rbrace}\label{eq:60}
\end{align}
The connection here, unlike for its Riemannian counterpart depends not only on the metric, but also on the non-metricity tensor $Q_{i}$. The arbitrary vector function $Q_{i}$  was perceived by Weyl as a linear 1-form, $d Q = {Q_{i}} d x^{i}$. Wherein ${\Gamma_{ij}}^{k}$ is invariant under gauge transformations. The exterior derivation of 1-form $dQ$, yields gauge invariant 2-form, $F = {F_{ij}} {d x^{i}}\wedge{d x^{j}}$, with ${F_{ij}} = {Q_{i,j}} - {Q_{j,i}}$ follows. This was called line curvature by Weyl by identifying $Q$ as the electromagnetic 4-vector potential he arrived at the Electromagnetic field tensor $F$.\\
Consider now the parallel transport of length, for instance; the norm of the tangent vector ($|X|$) along a curve $C$ with parameter $u$ to a different (but infinitesimally) neighbouring point:
\begin{align}
\frac{d |X|}{d u} = {{|X|}_{||k}} {X^{k}} = \left( \sigma - \frac{1}{2} {Q_{k}}{X^{k}}\right) {|X|}\label{eq:61}
\end{align}
The proper choice of curve parameter can yield us to write \eqref{eq:61} as $d|X| = - {Q_{k}}{X^{k}}{|X|} d u$, integrating along $C$ to get $|X| = \int \text{exp}( - Q(x) d x^{k})$, if $X$ is tangent to $C$; ${X^{k}} = \frac{d x^{k}}{d u}$, then:
\begin{align}
\left|\frac{d x^{k}}{d u}\right| = \int \text{exp}( - Q(x) d x^{k})\label{eq:62}
\end{align}
The length of the vector is not integrable since its value is dependent on the curve along which it is being parallel transported. Same holds for angle between parallel transported tangent vectors cf.~\eqref{eq:43}. For vanishing Elecromagnetic field, 4-potential becomes a gradient ("pure gauge"), such that, ${Q_{k}}{d x^{k}} = \frac{\partial \omega}{\partial {x^{k}}} {d x^{k}} = d\omega$, which incurs an integral independent of curvature.\\
\begin{center}
\textit{Thus in Weyls connection, both electromagnetic and gravitational fields represented by the vector field $Q$ and metrical field $g$, are intertwined.}
\end{center}
Due to the additional group of gauge transformations; $\rightarrow$ we must introduce Gauge Weight as a concept to tensor calculus.\\
The Lagrangian density $\mathcal{L} = - \sqrt{g} L$ must have gauge weight $w=0$ and the scalar $L$ by consequence has a gauge weight $-2$.\\
The only possibilities which are quadratic in the curvature tensor and line curvature, are given as (by Weitzenb\"{o}ck):
\begin{align}
\left( {K_{ij} {g^{ij}}}\right)^{2}; \hspace{3mm} {K_{ij}}{K_{kl}}{g^{ik}}{g^{jl}}; \hspace{3mm} {{K^{i}}_{jkl}}{{K^{j}}_{imn}}{g^{km}}{g^{ln}}; \hspace{3mm} {F_{ij}}{F_{kl}}{g^{ik}}{g^{jl}}\label{eq:63}
\end{align}
The last contingency leads to Maxwells equations, from invariants quadratic in curvature, in general, field equations of the fourth order emerge.\\
Weyls curvature tensor, derived from his connection, as given by Sch\"{o}uten {from \eqref{eq:49}}:
\begin{align}
{{K^{i}}_{jkl}} = {{R^{i}}_{jkl}} + {Q_{j;[}}{\delta_{l]}} + {\delta^{i}_{j}}{Q_{[l;k]}} - {Q^{i}_{;[k}}{g_{l]j}} +\frac{1}{2} {\left( {- {\delta^{i}_{[l}} {Q_{k]}}{Q_{j}} + {Q_{[k}}{g_{l]j}}{Q^{i}} -{\delta^{i}_{[k}}{g_{l]j}}{Q_{r}}{Q^{r}}}\right)}\label{eq:64}
\end{align}
If the metric field $g$ and vector potential $Q_{i} \equiv A_{i}$ are varied, independently from each of the curvature-dependent scalar invariants, we get contribution to the Maxwells equations.\\
F\"{o}rster (who published under the nom de plume "Bach") went through the mathematics of Weyls theory, including the calculation of the curvature tensor, quadratic Lagrangian field equations, exact solutions, taking $\mathcal{L} = \sqrt{g}{(3{W_{4}} - 6{W_{3}} + {W_{2}})}$, with:
\begin{align}
{W_{4}} = {S_{pqik}}{S^{pqik}}, \hspace{1mm} {W_{3}} = {g^{ik}}{g^{lm}}{{F^{p}}_{ikq}}{{F^{q}}_{lmp}}, \hspace{1mm} {W_{2}} = {g^{ik}}{{F^{p}}_{ikq}} \label{eq:65}
\end{align}
Where:
\begin{align}
{S_{[pq][ik]}} = \frac{1}{4} ({F_{pqik}} - {F_{qpik}} + {F_{ikpq}} -{F_{kipq}})\label{eq:66}
\end{align}
And further:
\begin{equation}
\begin{aligned}
{F_{pqik}} = {R_{pqik}} + \frac{1}{2} ({g_{pq}}{f_{ki}} + {g_{pk}}{f_{q;i}} + &{g_{qi}}{f_{p;k}} - {g_{pi}}{f_{q;k}} - {g_{qk}}{f_{p;i}}) + \frac{1}{2} ({f_{q}}{g_{p;[k}}{f_{i]}} + \\ &{f_{p}}{g_{q;[k}}{f_{i]}} + {g_{p[i}}{g_{k]q}}{f_{r}}{f^{r}})\label{eq:67}
\end{aligned}
\end{equation}
${R_{pqik}}$ is the Riemannian curvature tensor, ${f_{ik}} = {f_{i,k}} - {f_{k,i}}$, $f_{i}$ being the Electromagnetic 4-potential.
\newpage
\section{The physical point of view}
Weyls unification of Electromagnetism and Gravitation, although splendid because of its mathematical elegance and beauty, was bereft of physical substance.\\
In Einsteins General Relativity, the line element $ds$, is identified with space-time intervals measurable by real clocks and real measurement rods.\\
Since in Weyls theory only the equivalence class $\lbrace \lambda {g_{ik}} | \lambda \hspace{0.85mm} {\text{arbitrary}}\rbrace$, was supposed to have physical meaning; the clocks and rulers could be arbitrarily "regauged" in each event; whilst in Einsteins theory, the same clocks and rulers are to be used everywhere.\\
\underline{Einstein argued:} in his “Addendum” (“Nachtrag”) to Weyl’s paper in the
reports of the Academy (because Nernst had insisted on such a postscript). There, Einstein argued
that if light rays were to be the only available means for the determination of metrical relations
near a point, then Weyl’s gauge would make sense. However, as long as measurements are made
with (infinitesimally small) rigid rulers and clocks, there is no indeterminacy in the metric (as
Weyl would have it): Proper time can be measured. As a consequence it follows: If in nature length
and time would depend on the pre-history of the measuring instrument, then no uniquely defined
frequencies of the spectral lines of a chemical element could exist, i.e., the frequencies would depend
on the location of the emitter. Is his own words:
\begin{center}
“Regrettably, the basic hypothesis of the theory seems unacceptable to me, [of a theory]
the depth and audacity of which must fill every reader with admiration.”
\end{center}
Path dependence of the frequencies of spectral lines stems from the path dependency of \eqref{eq:62}.\\
Only for a vanishing Electromagnetic field is this untrue.\\
\underline{H.A Lorentz:} in a paper on the measurement of length and time intervals in GR and generalisations, he contradicted Weyls statement that world lines of light signals would suffice to determine the gravitational potentials.
\section{The Second Clock effect}
A space-time model is said to demonstrate a "Second Clock Effect", if the clock rate of clocks depends on their histories. The deduction of whether a Weyl structure presents such an effect has been usually done on largely intuitive basis, using the fact that the norms of parallel transported vector along a time-like curve represents clock rate of a clock. Therein, the spectral lines emitted by the clock could depend on history, which hasn't been experimentally observed, so far. As quoted in \cite{Avalos2016unj}:\\
\begin{figure}[ht!]
\centering
\captionsetup{justification=centering,margin=2cm}
\includegraphics[scale=.85]{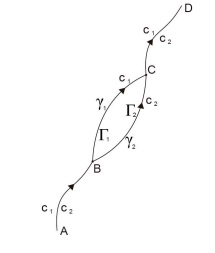}
\caption{Transport of two clocks $C_{1}$ and $C_{2}$. The curves $\gamma_{1}$ and $\gamma_{2}$, represent worldlines of $C_{1}$ and $C_{2}$, respectively, whilst $\Gamma_{1}$ and $\Gamma_{2}$ denote the portions of these curves between $B$ and $C$ \protect\footnotemark}
\label{fig:SCE}
\end{figure}
\\
In order to give an answer to this question consider the following situation. Suppose that
we transport two identical standard clocks along a time-like curve from $A$ to $B$, and then, at
$B$, they separate, following different paths $\gamma_{1}$ and $\gamma_{2}$ until they merge again at event $C$ (See \ref{fig:SCE}), after which they continue their journey together along the same path. Suppose that
both clocks were synchronized at $A$. Thinking of a clock as a device that counts the number
of cycles of some periodic process, what we are saying when we refer to synchronization is
that identical clocks use the same type of process and that the periods of these cycles were
set to be equal at $A$ (both clocks are set with the same scale at $A$).
Now assume that our space-time model does not exhibit a second clock effect. Accepting this hypothesis means
that we would expect that the clock rate of a clock, at a given event in space-time, should
depend only on local properties of the clock, that is, its position, instant velocity, instant
acceleration, etc, but not on its history. Therefore, in the particular case we are considering,
after we bring back the two identical clocks together at $C$, and keep them together, we would
expect their clock rates to coincide. In other words, we would expect the number of cycles
counted by either clock after $C$ to be the same. This also means that the readings of the
two clocks would coincide at any subsequent event D ($\tau_{\text{CD}}$ = ${\xoverline{\tau}}_{\text{CD}}$). These considerations
give us a way to test a possible existence of the second clock effect: compute the elapsed
time for both clocks between $C$ and some subsequent event $D$ and see whether they agree
or not. If they do not, then clearly there is a second clock effect.
\footnotetext{image reference: \cite{LOBO2018306}}
\section{Reactions to Weyls Theory}
\begin{enumerate}[label = \Roman*.]
\item {\underline{Einstein and Weyl:}
        \begin{itemize}[label = $\square$]
         \item Einstein despite liking the mathematical elegance of Weyls theory, persisted in denying its applicability to nature.
         \item Einstein liked Weyls argument in support of 4-dimensional space-time. As the Lagrangian for the Electromagnetic field is of gauge weight $-2$, $\sqrt{-g}$ has gauge weight $\frac{D}{2}$ in $M_{D}$, the integrand in the Hamiltonian principle can weigh zero only for $D=4$.
         \item After Weyls paper on "pure infinitesimal geometry" had been published, Einstein posed further doubts about Weyls theory: 
            \begin{enumerate}
                \item Weyls theory preserves similarity of geometric figures under parallel transport, but this would not be the most general situation.
                \item Suggested the affine group as the more general setting for generalisation of Riemannian geometry.
            \end{enumerate}
          \item According to Einstein, the line element '$ds$', was no longer a measurable quantity, in the premise of Weyls theory and the Electromagnetic 4-potential had never been one and was ergo, not a substitute.
          \item Weyl, in order to refute Einsteins claims, presented a new argument:
                \begin{itemize}
                    \item The quadratic form ${R}{g_{ik}}{d x^{i}}{d x^{k}}$ is an absolute invariant, i.e. with regards to gauge transformations as well (with gauge weight 0) . If this is taken to be the measurable distance in place of $ds$, then:
                        \begin{center}
                         "By prefixing this factor, (the) absolute norming of the unit of length is achieved, after all."
                        \end{center}
                \end{itemize}
           \item  Einstein thought that this was well off-colour, since $R$ was greatly dependent on matter density, ergo a small change in measuring path profoundly influences the integral of the square of the quantity.
           \item This in itself, isn't a wholly convincing counter-argument, since $g_{ik}$ heavily relies on matter density via his field equations, as well.
           \item In view of the more general Quadratic Lagrangian needed in Weyls theory, the connection between $R$ and the matter tensor might become less direct.
        \end{itemize}
\item \underline{Sch\"{o}uten, Pauli, Eddington and others:}
        \begin{itemize}[label = $\square$]
         \item \underline{Sommerfeld} seems to have agreed with Weyls theory and thought it was brilliant, likening it to Mies attempt \cite{Corry2018}.
         \item \underline{Sch\"{o}uten} proved that Weyls connection was indeed gauge invariant as well as making the identification of the 2-form $F_{\mu\nu}$ with the electromagnetic 4-potential, making no comments about the physics in question.
         \item \underline{Pauli} described the basic elements of Weyls theory in an article for the Encyclopaedia of mathematical sciences, talking about:
            \begin{itemize}
                \item Basic elements of geometry
                \item loss of line element $ds$ as a physical variable
                \item A convincing derivation for the conversation law of electric charge
                \item Too many possibilities for a Lagrangian inherent in a homogenous function of degree 1 of the invariants \eqref{eq:63}
                \item how no progress was made in describing the constituents of matter, since:
                        \begin{itemize}
                         \item differential equations too complicated to be solved
                         \item mass difference between elementary particles with positive and negative charge remained unexplained.
                        \end{itemize}
                \item This criticism was not limited to Weyls theory, albeit for all continuum theories and those which treat the electron or other constituent particles a singularities.
                \item Einstein though that one must pass to tensors of the fourth order, rather than only to those of second order, which carries with it, a vast indeterminacy, because:
                    \begin{enumerate}
                        \item Many more equations to take into account
                        \item Solutions contain more arbitrary constants
                    \end{enumerate}
                \item \underline{Eddington} gave a non-technical introduction to Weyls Theory (in his book \cite{eddington_2013}). Wherein the idea of gauging lengths independently was the central theme.
                While fourfold freedom in choice of coordinates gives the laws of conversation of energy and momentum; the new geometry thus has a fifth arbitrariness due to that in choosing the gauge, which further gives rise to a novel identity which represents the conversation law of electric charge.
            \end{itemize}
        \item He distinguished natural geometry and actual space from world geometry and conceptual space which can be contrived as graphical representation of relationships among physical observables.
        \item He further believed that Weyl was approaching the idea from the wrong end, i.e:
                \begin{center}
                \justify
                 Weyls non-Riemannian geometry was not to be applied to actual spacetime; albeit to a graphical representation of the relation-structure which is the basis of all physics, and both electromagnetic and metrical variables appear in it as interrelated.
                \end{center}
        \end{itemize}
}
\end{enumerate}
\section{Further Research}
\begin{itemize}[label = $\square$]
 \item \underline{Pauli} Weyls theory had, for the static case with a possible solution manifesting as a constant Ricci scalar and also admitted the Schwarzschild solution and could reproduce desired effects.
 \item \underline{Weyl} returned to the idea of gauging length by setting $R = \lambda = \text{constant}$, he thought of $\lambda$ as "Radius of curvature" of the world. (Again refer to \cite{eddington_2013} for more on the idea of "Radius of curvature of the world").
 \item In 1919, Weyls Lagrangian originally was: $\mathcal{L} = \frac{1}{2} {\sqrt{g}}{K^{2}} + {\beta}{F_{ik}}{F^{ik}}$ with the constraint that $K = 2\lambda$, where $\lambda = \text{constant}$, as an equivalent Lagrangian, Weyl gave up to a divergence:
\begin{align}
\mathcal{L} = {\sqrt{g}}(R + {\alpha}{F_{ik}}{F^{ik}} + \frac{1}{4}\left(2\lambda - 3{\phi_{l}}{\phi^{l}})\right)\label{eq:68}
\end{align}
With 4-potential $\phi_{l}$ and electromagnetic field $F_{ik}$. Due to the constraint imposed, the problem of formulation of Cauchy initial value problem for field equations of fourth order had been circumvented by a reduction  to second order field equations.
\item Later in a 1921 paper he changed his Lagrangian slightly to:
\begin{align}
\mathcal{L} = {\sqrt{g}}(R + {\alpha}{F_{ik}}{F^{ik}} + \frac{\epsilon}{4}\left(1 - 3{\phi_{l}}{\phi^{l}})\right)\label{eq:69}
\end{align}
$\epsilon$ being a factor in Weyls connection:
\begin{align}
{\Gamma_{ij}}^{k} =  \stirlingii{k}{i \hspace{2mm} k} + \frac{\epsilon}{2}({\delta^{k}_{i}}{\phi_{j}} + {\delta^{k}_{j}}{\phi_{i}} - {g_{ij}}{g^{kl}}{\phi_{l}})\label{eq:70}
\end{align}
An advantage according to him was that his theory led to the cosmological term $\lambda$ in a uniform and forceful manner, which in Einsteins theory was adhoc.
\item \underline{Reichenb\"{a}cher} was unhappy with Weyl taking curvature to be constant before the variation and wondered if introducing Weyls natural gauge after the variation of the Lagrangian such that the field equations could show gauge invariance first.
\item \underline{Eddington} thought Weyls choice of Lagrangian was highly speculative.
\end{itemize}
This brings us to understand the manner in which Weyls theory came into being and how it was received by the community back then. We can now begin to lay the groundwork to discuss the double layers themselves.

\chapter{Double Layer Theory}
\label{chap:DLT}
\section{Definitions}
\label{sec:defn}
Having dealt with the geometrical preliminaries and having introduced Weyl Gravity, we can proceed to establish the intricacies of Double Layer Theory. We start with the Weyl action which we define as:
\begin{align}
\mathbb{S}_{W} = \int \mathcal{L}_{W}\sqrt{-g}\,d^4 x\label{eq:71}
\end{align}
The Quadratic Lagrangian is taken to be:
\begin{align}
\mathcal{L}_{W} = \alpha_1 {{R^{\mu}}_{\nu\lambda\sigma}}{{R_{\mu}}^{\nu\lambda\sigma}} + \alpha_2 {R^{\mu\nu}}{R_{\mu\nu}} + \alpha_3 {R^2} + \alpha_4 {F^{\mu\nu}}{F_{\mu\nu}}\label{eq:72}
\end{align}
The definitions used for Curvature Tensor, the Ricci Tensor and the curvature scalar, respectively, are:
\begin{align}
 {{R^{\mu}}_{\nu\lambda\sigma}} = {\frac{\partial {\Gamma^\mu_{\nu\sigma}}}{\partial X^\lambda}} - {\frac{\partial {\Gamma^\mu_{\nu\lambda}}}{\partial X^\sigma}} + {\Gamma^\mu_{\kappa\sigma}}{\Gamma^\kappa_{\nu\lambda}} - {\Gamma^\mu_{\kappa\lambda}}{\Gamma^\kappa_{\sigma\nu}}\label{eq:73}
\end{align}
\begin{align}
 {{R}_{\mu\nu}} = {\frac{\partial {\Gamma^\lambda_{\mu\nu}}}{\partial X^\lambda}} - {\frac{\partial {\Gamma^\lambda_{\nu\lambda}}}{\partial X^\nu}} + {\Gamma^\lambda_{\kappa\nu}}{\Gamma^\kappa_{\mu\lambda}} - {\Gamma^\lambda_{\kappa\lambda}}{\Gamma^\kappa_{\mu\nu}}\label{eq:74}
\end{align}
\begin{align}
R = {g^{\mu\nu}}{R_{\mu\nu}}\label{eq:75}
\end{align}
Here, ${\Gamma^\mu_{\nu\sigma}}$, is our composite connection, comprised of two parts, the Riemannian Connection (represented by the Christoffel Symbols ${C^\mu_{\nu\sigma}}$) and the Weylian connection (${W^\mu_{\nu\sigma}}$):
\begin{align}
{\Gamma^\mu_{\nu\sigma}} = {C^\mu_{\nu\sigma}} + {W^\mu_{\nu\sigma}}\label{eq:76}
\end{align}
These are further defined each as:
\begin{align}
{C^\mu_{\nu\sigma}} = \frac{1}{2} {{g^{\mu\kappa}}\left((g_{\kappa\nu})_{,\sigma} + (g_{\kappa\sigma})_{,\nu} - (g_{\sigma\nu})_{,\kappa}\right)}\label{eq:77}
\end{align}
\begin{align}
{W^\mu_{\nu\sigma}} = - \frac{1}{2} {\left({{A_\nu}{\delta^\mu_\sigma}} + {{A_\sigma}{\delta^\mu_\nu}} - {A^\mu}{g_{\nu\sigma}}\right)}\label{eq:78}
\end{align}
Further we define the two-form $F_{\mu\nu}$ as:
\begin{align}
F_{\mu\nu} = {\nabla_{\nu}}{A_\mu} - {\nabla_{\mu}}{A_\nu} = A_{\mu;\nu} - A_{\nu;\mu}\label{eq:79}
\end{align}
Here, $A_\mu$, is a one-form, which gauges the theory. it is defined in the non-metricity condition, which negates metric compatibility as we've seen in \ref{chap:intro}, with equation \cref{eq:1}. 
Non-metricity can be written down as:
\begin{align}
\nabla_\lambda g_{\mu\nu} = A_\lambda g_{\mu\nu} = Q_{\mu\nu\lambda} \label{eq:80}
\end{align}
$\nabla_\lambda$ being the covariant derivative with respect to the $\lambda$ index. ";" being the conventional method of representing it, with respect to the succeeding index. Whilst "," is the partial derivative.

\section{A description of the volume of integration}
\label{sec:vol}
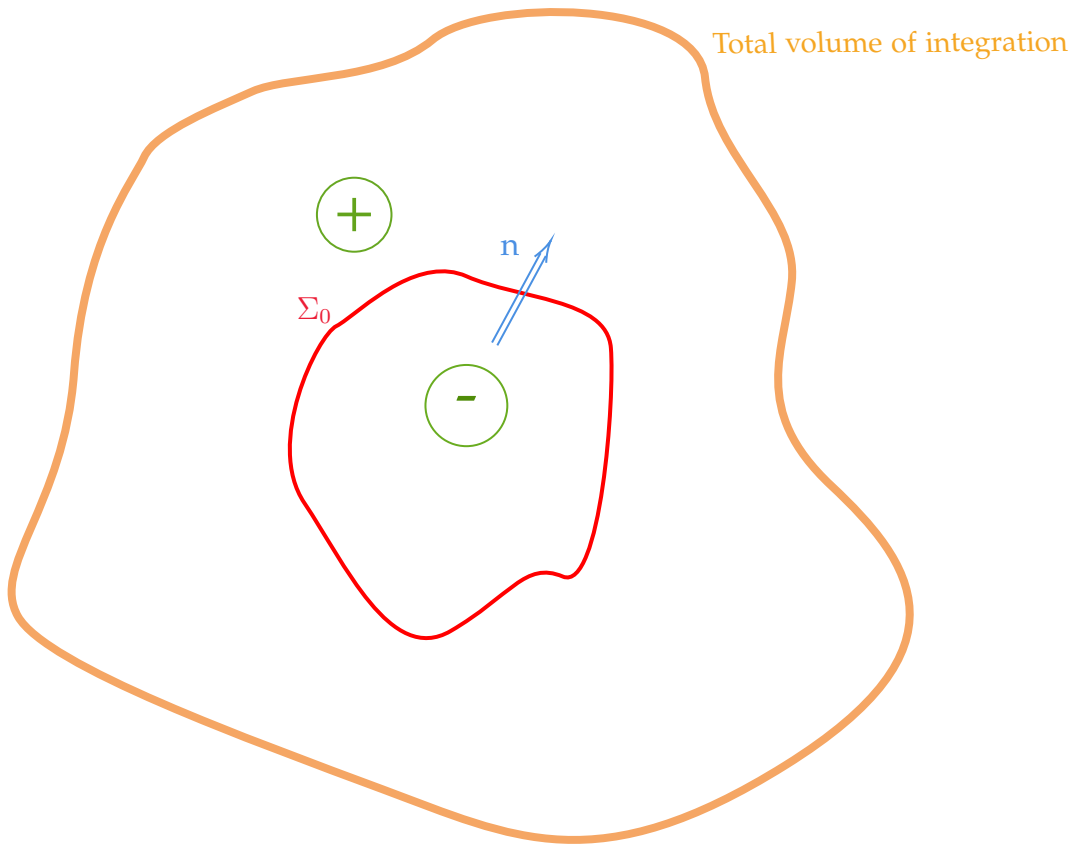
\begin{figure}[h!]
\captionsetup{justification=centering,margin=2cm}
\centering
 \tikzset{every picture/.style={line width=0.75pt}} 

\begin{tikzpicture}[x=0.75pt,y=0.75pt,yscale=-1,xscale=1]

\draw  [color={rgb, 255:red, 245; green, 166; blue, 100 }  ,draw opacity=1 ][line width=3]  (142.5,91) .. controls (148.5,78) and (178.51,66.14) .. (196.5,58) .. controls (214.49,49.86) and (261.5,53) .. (286.5,32) .. controls (311.5,11) and (418.5,11) .. (422.5,51) .. controls (426.5,91) and (469.3,118.46) .. (465.81,153.57) .. controls (462.31,188.69) and (444.57,217.66) .. (485.54,256.33) .. controls (526.5,295) and (556.5,338) .. (462.5,397) .. controls (368.5,456) and (327.14,433.63) .. (261.5,409) .. controls (195.86,384.37) and (93.5,347) .. (79.5,322) .. controls (65.5,297) and (102.5,267) .. (107.5,200) .. controls (112.5,133) and (136.5,104) .. (142.5,91) -- cycle ;
\draw  [color={rgb, 255:red, 255; green, 0; blue, 0 }  ,draw opacity=1 ][line width=1.5]  (238.5,176) .. controls (243.75,173.37) and (255.46,162.07) .. (269.34,154.79) .. controls (280.09,149.16) and (292.13,145.93) .. (303.5,151) .. controls (329.56,162.62) and (373.5,162) .. (375.5,187) .. controls (377.5,212) and (370.5,310) .. (351.5,302) .. controls (332.5,294) and (325.47,311.85) .. (295.48,329.42) .. controls (265.5,347) and (242.5,295) .. (222.5,265) .. controls (202.5,235) and (229.18,180.66) .. (238.5,176) -- cycle ;
\draw [color={rgb, 255:red, 74; green, 144; blue, 226 }  ,draw opacity=1 ][fill={rgb, 255:red, 74; green, 144; blue, 226 }  ,fill opacity=1 ]   (316.18,184.29) -- (340.39,139.33)(318.82,185.71) -- (343.03,140.75) ;
\draw [shift={(345.5,133)}, rotate = 478.3] [color={rgb, 255:red, 74; green, 144; blue, 226 }  ,draw opacity=1 ][line width=0.75]    (10.93,-3.29) .. controls (6.95,-1.4) and (3.31,-0.3) .. (0,0) .. controls (3.31,0.3) and (6.95,1.4) .. (10.93,3.29)   ;

\draw (424.58,26.56) node [anchor=north west][inner sep=0.75pt]  [font=\small,color={rgb, 255:red, 245; green, 166; blue, 35 }  ,opacity=1 ,rotate=-359.64,xslant=-0.04] [align=left] {Total volume of integration};
\draw (319,131) node [anchor=north west][inner sep=0.75pt]   [align=left] {\textcolor[rgb]{0.29,0.56,0.89}{n}};
\draw  [color={rgb, 255:red, 105; green, 172; blue, 32 }  ,draw opacity=1 ]  (303.38, 216) circle [x radius= 20.43, y radius= 20.43]   ;
\draw (303.38,216) node  [font=\fontsize{0.7em}{0.84em}\selectfont,color={rgb, 255:red, 101; green, 167; blue, 29 }  ,opacity=1 ] [align=left] {\textcolor[rgb]{0.31,0.55,0.03}{{\Huge -}}};
\draw  [color={rgb, 255:red, 103; green, 167; blue, 34 }  ,draw opacity=1 ]  (247.38, 120) circle [x radius= 18.61, y radius= 18.61]   ;
\draw (247.38,120) node  [font=\fontsize{0.7em}{0.84em}\selectfont] [align=left] {\textcolor[rgb]{0.39,0.64,0.11}{{\huge +}}};
\draw (227.55,168.05) node  [color={rgb, 255:red, 236; green, 29; blue, 54 }  ,opacity=1 ] [align=left] {$\displaystyle \Sigma _{0}$};

\end{tikzpicture}
\caption{A diagram representing the volume of integration}\label{fig:vol}
\end{figure}
 
Here, the orange outline represents the boundary of the total volume across which we integrate the Weylian action. The red outline represents our \textit{singular hypersurface} $\Sigma_0$, which is detailed in Israel's seminal 1966 paper on Singular Hypersurfaces and Thin shells in GR \cite{Israel1966}. This singular hypersurface separates the volume of integration into two regions as visible, the "$-$" and the "$+$" regions, with an outward pointing normal $n(X^\mu)$ as shown (from $n<0$ to $n>0$). This hypersurface, $\Sigma_0$, is singular in the sense that the extrinsic curvature in its vicinity in the $-$ and $+$ regions are not equal ($K^{-}_{ij}\neq K^{+}_{ij}$), further the normal vanishes on the surface $n(X^\mu) = 0$.\\
We can now discuss the total action and what it is comprised of. The total action $S_{\text{tot}}$ can be decomposed into its Weylian part $S_{W}$ and the Matter part $S_{M}$:
\begin{align}
S_{\text{tot}} = S_{W} + S_{M}\label{eq:81}
\end{align}
Let's look at the variation of the matter action first: \
\begin{align}
\delta S_{M} = -\frac{1}{2} \int {T^{\mu\nu}} (\delta g_{\mu\nu})\, \sqrt{-g}\, {d^4 x} - \int G^\mu (\delta A_\mu)\, \sqrt{-g}\,d^4 x
\label{eq:82}
\end{align}
As is apparent, the matter action is made of its Surface energy tensor $-\frac{1}{2} \int {T^{\mu\nu}} (\delta g_{\mu\nu}) \sqrt{-g} {d^4 X}$ and "Charge" (not precisely analogous to electric charge) $\int G^\mu (\delta A_\mu) d^4 X$ components. $T^{\mu\nu}$ represents the surface energy-momentum tensor and $G^\mu$ represents "charge".\\
Since $\Sigma_0$ is singular, we can write down our surface and charge energy tensor and one-form as their decompositions in terms of the delta function (which itself is a function of the normal $n$) on the surface itself, with some contributions from the bulk, which does not fall under the purview of this report since we are only interested in the surface itself.we express this like so:
\begin{align}
T^{\mu\nu} = S^{\mu\nu} \delta(n) + \cdots \label{eq:83}
\end{align}
\begin{align}
G^{\mu} = Q^{\mu} \delta(n) + \cdots \label{eq:84}
\end{align}
Reintroducing these (\eqref{eq:83} and \eqref{eq:84}) into \eqref{eq:82}, we obtain:
\begin{align}
\delta S_{M}\big|_{\Sigma_0} = -\frac{1}{2} \int_{\Sigma_0} {S^{\mu\nu}} (\delta g_{\mu\nu})\, \sqrt{\lvert\gamma\rvert}\, {d^3 X} - \int_{\Sigma_0} Q^\mu (\delta A_\mu)\, \sqrt{\lvert\gamma\rvert}\, {d^3 X}\label{eq:85}
\end{align}
Where, $\gamma_{ij}$, is the three dimensional metric bound to $\Sigma_0$, the interval is as given:
\begin{align}
d s^2\vert_{\Sigma_0} = {\gamma_{ij}} d X^{i} d X^{j} \hspace{3mm} \label{eq:86}
\end{align}
Where, $x^{i} \in \Sigma_0$.\\
With this description complete, we move on to next description of the Gauss Normal Coordinate System.
\section{Gauss Normal Coordinate System}
\label{sec:gncs}
We shall be using the Gauss normal coordinate system since it will help us in simplifying calculations involving jumps/discontinuities across $\Sigma_0$ a ways down the line. For our case, we define the generalised form as follows:
\begin{align}
X^\mu = \lbrace n,X^i\rbrace\label{eq:87}
\end{align}
\begin{align}
d s^2 &= {g_{ij}}{dX^i}{dX^j}\\
d s^2 &= {g_{nn}}{d n^2} + 2{g_{ni}}{dn}{dX^i} + {g_{ij}}{d X^{i}}{d X^{j}}\label{eq:88}
\end{align}
Where the normal components of the four-dimensional metric are clearly segregated from the others.
Further, we can define our surface metric in terms of the four dimensional metric in the Gauss normal coordinate system, like so:
\begin{align}
\gamma_{ij} = g_{ij} - \frac{{g_{ni}}{g_{nj}}}{g_{nn}}\label{eq:89}
\end{align}
With regards to our hypersurface $\Sigma_0$, we have:
\begin{align}
{g_{nn}} &= \epsilon = \pm 1\\
{g_{ni}} &= 0
\label{eq:90}
\end{align}
Where the scalar "$\epsilon$" is positive or negative depending on whether we are in a timelike or a spacelike spacetime. Despite \eqref{eq:90}, the variations in $g_{nn}$ and $g_{ni}$ do not vanish, i.e. $\delta g_{nn} \neq 0$ and $\delta g_{ni} \neq 0$. Further the variation in the surface metric is written as:
\begin{align}
\delta\gamma_{ij} = \delta {g_{ij}} - \frac{g_{ni}}{g_{nn}}\delta{g_{nj}} - \frac{g_{nj}}{g_{nn}}\delta{g_{ni}} + \frac{{g_{ni}}{g_{nj}}}{({g_{nn}})^2}\delta {g_{nn}}\label{eq:91}
\end{align}
This variation is reduced considerably in the Gauss normal system:
\begin{align}
\left(\delta{\gamma_{ij}}\right)_{,n} = \left(\delta{g_{ij}}\right)_{,n}
\label{eq:92}
\end{align}
Using the Gauss normal coordinate system as our base and with the knowledge of the the normal components of the metric at hand, we can calculate the Christoffel symbols using \eqref{eq:77}, which come out to be:
\begin{align}
{C^n_{nn}} = {C^n_{ni}} = 0;\hspace{3mm} {C^n_{ij}} = \epsilon {K_{ij}};\hspace{3mm} {C^i_{nj}} = -{K^i_j}\label{eq:93}
\end{align}
$K_{ij}$ is the extrinsic curvature of $\Sigma_0$ as:
\begin{align}
{k_{ij}} = -\frac{1}{2}\frac{\partial {g_{ij}}}{\partial n} = -\frac{1}{2}({g_{ij,n}}) \hspace{3mm} (\text{on}\hspace{0.85mm}{\Sigma_0})
\label{eq:ext}
\end{align}
Now that we have all these quantities at our disposal, we can use \eqref{eq:77} and \eqref{eq:78} to calculate the variations in the Christoffel Symbols and the Weylian connection:
\begin{align}
\delta{C^\lambda_{\mu\nu}} = \frac{1}{2}{g^{\lambda\kappa}}\left((\delta{g_{\kappa\mu}})_{;\nu} + (\delta{g_{\kappa\nu}})_{;\mu}\right) - (\delta{g_{\mu\nu}})_{;\kappa}\label{eq:94}
\end{align}
The variation of the Weylian Connection is a bit more rigorous and extensive as we see:
\begin{align}
\delta W^\lambda_{\mu\nu} = -\frac{1}{2}\left((\delta A_{\mu}){\delta^\lambda_\nu} + (\delta A_{\nu}){\delta^\lambda_\mu} - {A^\lambda}(\delta {g_{\mu\nu}}) - {g_{\mu\nu}}(\delta {A^\lambda})\right)\label{eq:95}
\end{align}
Further, $\delta A^\lambda$is expressed as:
\begin{align*}
A^\lambda &= g^{\lambda\kappa} {A_{\kappa}}\\
\delta A^\lambda &= g^{\lambda\kappa} \left(\delta {A_{\kappa}}\right) + {A_{\kappa}} \left(\delta g^{\lambda\kappa}\right)\\
\delta g^{\lambda\kappa} &= - {g^{\lambda\alpha}}{g^{\kappa\rho}}\left(\delta g_{\alpha\rho}\right)
\end{align*}
Finally, we also calculate the variation in the Curvature tensor and Ricci tensor (using Palatini's 1919 formalism):
\begin{align}
\delta {{R^{\mu}}_{\nu\lambda\sigma}} &= {\nabla_\lambda}\left(\delta {\Gamma^\mu_{\nu\sigma}}\right) - {\nabla_\sigma}\left(\delta {\Gamma^\mu_{\nu\lambda}}\right) \hspace{5mm}{(\textbf{Curvature Tensor Variation})}\\
\delta {R^{\mu\nu}} &= {\nabla_\lambda}\left(\delta {\Gamma^\lambda_{\nu\sigma}}\right) - {\nabla_\sigma}\left(\delta {\Gamma^\lambda_{\nu\lambda}}\right) \hspace{4mm}(\textbf{Ricci Tensor Variation})\label{eq:96}
\end{align}
We can now commence the variation process, which we do in the next section.
\newpage
\section{The variation scheme described}
\label{sec:plan}
Solving the action integral involves, as a quintessential part, to perform the variation of the various quantities under integration, albeit selectively, as we will learn later \ref{subsubsec:detour}, we need only the essential variation coefficients which contribute meaningfully to the Double Layer. Simplifying this process of solving various discontinuities and collecting the variation coefficient terms is rather extensive and we split the equation into it's four distinct comprising coefficients ($\alpha_1, \alpha_2, \alpha_3, \alpha_4$). The $\alpha_4$ case is relatively the least difficult one and hence we tackle it in the beginning as you'll see in \ref{subsec:a4}. The rest of them, despite being lengthy share certain steps which we apply to all of them:
\begin{enumerate}[label =\roman*.]
 \item Reducing the integral to a simpler form by utilising the Palatini formalism \eqref{eq:96}.
 \item Forming the total derivative including the curvature (tensor/scalar)/Ricci tensor and the $\Gamma$ variation using the covariant derivative and keeping aside the surface terms and any negation which might have been produced in the process.
 \item Implementation of the Stokes theorem for the first time to obtain discontinuities across our hypersurface, in terms of the surface metric and 3-dimensional Gauss normal coordinates.
 \item After getting the 4-dimensional discontinuities, we implement our normal symmetry and transition to the 3-dimensional discontinuities. 
 \item We then address the negations and surface integral terms and carry out the second iteration of the Stokes theorem.
 \item We can now collect all the relevant variation coefficients and explore their implications.
\end{enumerate}
Briskly equipped with our strategy, we head on to perform these steps one by one.
\section{The Variation}
\label{sec:var} 
Let's write down the integral for the variation in the Weyl Action:
\begin{align}
\delta {S_{W}} = \int \left(\delta {{\mathcal{L}}_{W}} - \frac{1}{2}{\mathcal{L}}\left(\delta g^{\mu\nu}\right)\right)\,\sqrt{-g}\,d^4 x\label{eq:97}
\end{align}
We can omit all terms with variation in the metric and the vector/one-form "potential", ${g_{\mu\nu}}\delta {g^{\mu\nu}}$ and $\delta {A_{\mu}}$, since they don not contain the delta function, which is a double layer characteristic. This is because $\delta S(\pm) = 0$ in the bulk on both sides of the surface.\\
Hence our variation integral reduces to:
\begin{multline}
\delta {S_{W}} \rightarrow \int \delta {{\mathcal{L}}_{W}}\,\sqrt{-g}\,d^4 x = \\2\int \lbrace \alpha_1 {R_\mu}^{\nu\lambda\sigma}({\delta {{R^{\mu}}_{\nu\lambda\sigma}}}) + \alpha_2 {R^{\mu\nu}}({\delta {R_{\mu\nu}}}) + \alpha_3 R{g^{\mu\nu}}(\delta R_{\mu\nu}) +\alpha_4 {F^{\mu\nu}}({\delta {F_{\mu\nu}}})\rbrace + ()(\delta {g_{\mu\nu}}) +()(\delta {A_\mu})
\label{eq:98}
\end{multline}
Which reduces to (due to omission of the ${g_{\mu\nu}}\delta {g^{\mu\nu}}$ and $\delta {A_{\mu}}$ terms):
\begin{multline}
\delta {S_{W}} \rightarrow \int \delta {{\mathcal{L}}_{W}}\,\sqrt{-g}\,d^4 x = \\2\int \lbrace \alpha_1 {R_\mu}^{\nu\lambda\sigma}({\delta {{R^{\mu}}_{\nu\lambda\sigma}}}) + \alpha_2 {R^{\mu\nu}}({\delta {R_{\mu\nu}}}) + \alpha_3 R(\delta R) +\alpha_4 {F^{\mu\nu}}({\delta {F_{\mu\nu}}})\rbrace\,\sqrt{-g}\,d^4 x
\label{eq:99}
\end{multline}
We now begin each variation, which I will be referring to here, using their coefficients (say $\alpha_1$ or $\alpha_2$, so on and so forth).\\
Let's commence with $\alpha_4$;
\subsection{Variation of the \texorpdfstring{$\alpha_4$}{alpha 4} term}
\label{subsec:a4}
The term in question is $\alpha_4 {F^{\mu\nu}\delta {F_{\mu\nu}}}$, we know from \eqref{eq:79}, that: ${F_{\mu\nu}} = {A_{\nu ; \mu}} - {A_{\mu ; \nu}} = {A_{\nu , \mu}} - {A_{\mu , \nu}} \hspace{3mm}(\text{Since terms with the connection cancel out in this difference})$, let us now try to write down the variation in this term, integrand only:
\begin{align}
{F^{\mu\nu}\delta {F_{\mu\nu}}} = {F^{\mu\nu}}{\left((\delta A_\nu)_{,\mu} - (\delta A_\mu)_{,\nu}\right)}\label{eq:100}
\end{align}
Putting this into the variation integral just for the $\alpha_4$ part:
\begin{align}
2{\alpha_4}\int{{F^{\mu\nu}}{\left((\delta A_\nu)_{;\mu} - (\delta A_\mu)_{;\nu}\right)}}\,\sqrt{-g}\,d^4 x\label{eq:101}
\end{align}
\begin{align}
4{\alpha_4}\int{F^{\mu\nu}}(\delta A_\nu)_{;\mu}\,\sqrt{-g}\,d^4 x \label{eq:102}
\end{align}
The difference in the variation of the $A_\mu$s is antisymmetric, yielding us twice the first term, with the covariant derivative with respect to $\mu$ index, (since the second term cancels out).\\ 
\eqref{eq:102} can be written as a total derivative with the extra portion ${{F^{\mu\nu}}_{;\mu}\delta {A_{\mu}}}$, as such:
\begin{align}
4{\alpha_4}\int\left({F^{\mu\nu}}\delta A_\nu\right)_{;\mu} - {({F^{\mu\nu}})_{;\mu}\delta {A_{\mu}}}\,\sqrt{-g}\,d^4 x\label{eq:103}
\end{align}
We focus on the total derivative first and compute its jump across $\Sigma_0$, in order to do this, we must implement Stokes principle, let's see how we can do this:
\begin{align}
4 \alpha_4 \int ({F^{\mu\nu}}\delta {A_{\nu}}\sqrt{-g})_{,\mu}\,d^4 X\label{eq:104}
\end{align}
This result can be obtained using:
\begin{align*}
{{A^\mu}_{;\mu}} = \frac{{({A^\mu\sqrt{-g}})}_{,\mu}}{\sqrt{-g}}
\end{align*}
\eqref{eq:104} further gives us:
\begin{align}
4\alpha_4\int ({F^{\mu\nu}}\delta{A_{\nu}}\sqrt{-g})_{,\mu}\,d^4 x\label{eq:105}
\end{align}
As a jump:
\begin{align}
-4\alpha_4\int [F^{\mu\nu}]\delta{A_\nu}\,\sqrt{-g}\,d^4 x\label{eq:106}
\end{align}
The negative sign in front of $4\alpha_4$ comes from the opposing directions of the normal and the jump being taken across two regions $+ (n>0)$ and $- (n<0)$. Along the surface using three dimensional Gauss normal coordinates;
\begin{align}
-4\alpha_4\int_{\Sigma_0} [F^{ni}]\delta{A_i}\,\sqrt{\lvert\gamma\rvert}\,d^3 X\label{eq:alpha1}
\end{align}
Now we have the easiest piece of the puzzle, with this knowledge at hand, we can further find the jumps for the rest of the pieces.
\subsection{Variation of the \texorpdfstring{$\alpha_1$}{alpha 1} term}
\label{subsec:al1}
The $\alpha_1$ is coefficient to the variation integral of the main, un-contracted curvature tensor, like so:
\begin{align}
2\alpha_1\int{{R_{\mu}}^{\nu\lambda\sigma}}{\delta {{R^{\mu}}_{\nu\lambda\sigma}}}\,\sqrt{-g}\,d^4 x\label{eq:107}
\end{align}
Here, we can now use the Palatini formalism we introduced in \eqref{eq:96} and proceed further:
\begin{align}
2\alpha_1\int{{R_{\mu}}^{\nu\lambda\sigma}}({\nabla_\lambda}\left(\delta {\Gamma^\mu_{\nu\sigma}}\right) - {\nabla_\sigma}\left(\delta {\Gamma^\mu_{\nu\lambda}}\right))\,\sqrt{-g}\,d^4 x\label{eq:108}
\end{align}
Using the antisymmetricity of the final two indices of the curvature tensor, we can further write this as (also similar to the $\alpha_1$ case):
\begin{align}
4\alpha_1\int{{R_{\mu}}^{\nu\lambda\sigma}}{\nabla_\lambda}\left(\delta {\Gamma^\mu_{\nu\sigma}}\right)\,\sqrt{-g}\,d^4 x\label{eq:109}
\end{align}
As we did in the previous subsection for $F^{\mu\nu}$, we can convert this into a difference between the total derivative and the negation:
\begin{align}
4\alpha_1\int\lbrace{\nabla_\lambda}\left({{R_{\mu}}^{\nu\lambda\sigma}}{\delta {\Gamma^\mu_{\nu\sigma}}}\right) - {\nabla_\lambda}\left({{R_{\mu}}^{\nu\lambda\sigma}}\right){\delta {\Gamma^\mu_{\nu\sigma}}}\rbrace\,\sqrt{-g}\,d^4 x\label{eq:110}
\end{align}
A few interesting calculation devices can be introduced here, which will help us to simplify our calculations by quite a fair bit, these will also be used in the latter parts of the variation, without explicitly mentioning them over and over again, the reader is advised to refer back to this point and implement them much the same as they are here. Let's take our total derivative term and represent it using a one-form, as such (since all other indices are being contracted across except for $\lambda$ on the top):
\begin{align}
{{R_{\mu}}^{\nu\lambda\sigma}}{\delta {\Gamma^\mu_{\nu\sigma}}} = {L^{\lambda}}\label{eq:111}
\end{align}
Ergo, the total derivative is represented as:
\begin{align}
{\nabla_\lambda}{L^{\lambda}} = {{L^{\lambda}}_{;\lambda}} + {W^\lambda_{\lambda\kappa}}{L^{\kappa}}\label{eq:112}
\end{align}
Now, the Weylian connection for \eqref{eq:112}, is calculated as:
\begin{align}
{W^\lambda_{\lambda\kappa}} &= -\frac{1}{2}{\left({{A_{\lambda}}{\delta^\lambda_\kappa}} + {{A_{\kappa}}{\delta^\lambda_\lambda}}- {{A^{\lambda}}{g_{\lambda\kappa}}}\right)}\\
&= -\frac{1}{2}(4{A_{\kappa}}) = -2{A_{\kappa}}\\
\Rightarrow {\nabla_\lambda}{L^{\lambda}} &= {{L^{\lambda}}_{;\lambda}} - 2{A_{\kappa}}{L^{\lambda}}
\label{eq:113}
\end{align}
We shall keep this computation on the side for a while and refer to it a tad later.\\
Another thing we need to make sure is that the $\delta$ functions disappear and do not persist through the calculations in places they did not exist before, to do this, we decompose the curvature tensor in terms of its $+$ and $-$ components on each bulk side of the surface, this is coupled with the Heaviside function:
\begin{align}
{{R_{\mu}}^{\nu\lambda\sigma}} = {{R_{\mu}}^{\nu\lambda\sigma}}(+)\Theta(n) + {{R_{\mu}}^{\nu\lambda\sigma}}(-)\Theta(-n)\label{eq:114}
\end{align}
Putting this form into \eqref{eq:110}, we obtain the following:
\begin{multline}
4\alpha_1\int\lbrace{\nabla_\lambda}\left(\left({{R_{\mu}}^{\nu\lambda\sigma}}(+)\Theta(n) + {{R_{\mu}}^{\nu\lambda\sigma}}(-)\Theta(-n)\right)(\delta {\Gamma^\mu_{\nu\lambda}})\right)\\ - {\nabla_\lambda}\left({{R_{\mu}}^{\nu\lambda\sigma}}(+)\Theta(n) + {{R_{\mu}}^{\nu\lambda\sigma}}(-)\Theta(-n)\right)(\delta {\Gamma^\mu_{\nu\lambda}})\rbrace
\label{eq:115}
\end{multline}
We take two cases ($+$ and $-$) separately.\\
$\circ$ Taking the $+ (n>0)$ case, first:
\begin{multline}
4\alpha_1\int\lbrace{\nabla_\lambda}\left({{R_{\mu}}^{\nu\lambda\sigma}}(+)(\delta {\Gamma^\mu_{\nu\lambda}})\right)(\Theta(n)) + ({{R_{\mu}}^{\nu\lambda\sigma}}(+))(\delta {\Gamma^\mu_{\nu\lambda}})({\nabla_\lambda} \Theta(n))\\ - {\nabla_\lambda}({{R_{\mu}}^{\nu\lambda\sigma}}(+))(\delta {\Gamma^\mu_{\nu\lambda}})(\Theta(n)) - ({{R_{\mu}}^{\nu\lambda\sigma}}(+))(\delta {\Gamma^\mu_{\nu\lambda}})({\nabla_\lambda}\Theta(n))\rbrace\label{eq:116}
\end{multline}
The derivative of the Heaviside function $\nabla_\lambda\Theta(n)$ is none other than the delta function $\delta(n){n_{,\lambda}}$.
\begin{multline}
4\alpha_1\int\lbrace{\nabla_\lambda}\left({{R_{\mu}}^{\nu\lambda\sigma}}(+)(\delta {\Gamma^\mu_{\nu\lambda}})\right)(\Theta(n)) + ({{R_{\mu}}^{\nu\lambda\sigma}}(+))(\delta {\Gamma^\mu_{\nu\lambda}})(\delta(n){n_{,\lambda}})\\ - {\nabla_\lambda}({{R_{\mu}}^{\nu\lambda\sigma}}(+))(\delta {\Gamma^\mu_{\nu\lambda}})(\Theta(n)) - ({{R_{\mu}}^{\nu\lambda\sigma}}(+))(\delta {\Gamma^\mu_{\nu\lambda}})(\delta(n){n_{,\lambda}})\rbrace\label{eq:117}
\end{multline}
And finally, we get:
\begin{align}
4\alpha_1\int\lbrace{\nabla_\lambda}\left({{R_{\mu}}^{\nu\lambda\sigma}}(+)(\delta {\Gamma^\mu_{\nu\lambda}})\right)(\Theta(n)) - {\nabla_\lambda}({{R_{\mu}}^{\nu\lambda\sigma}}(+))(\delta {\Gamma^\mu_{\nu\lambda}})(\Theta(n))\rbrace\label{eq:118}
\end{align}
$\circ$ Let's take the $- (n<0)$ case:
\begin{multline}
4\alpha_1\int\lbrace{\nabla_\lambda}\left({{R_{\mu}}^{\nu\lambda\sigma}}(-)(\delta {\Gamma^\mu_{\nu\lambda}})\right)(\Theta(-n)) + ({{R_{\mu}}^{\nu\lambda\sigma}}(-))(\delta {\Gamma^\mu_{\nu\lambda}})({\nabla_\lambda} \Theta(-n))\\ - {\nabla_\lambda}({{R_{\mu}}^{\nu\lambda\sigma}}(-))(\delta {\Gamma^\mu_{\nu\lambda}})(\Theta(-n)) - ({{R_{\mu}}^{\nu\lambda\sigma}}(-))(\delta {\Gamma^\mu_{\nu\lambda}})({\nabla_\lambda}\Theta(-n))\rbrace\label{eq:119}
\end{multline}
The derivative of the Heaviside function $\nabla_\lambda\Theta(-n)$ is none other than the delta function $-\delta(n){n_{,\lambda}}$.
\begin{multline}
4\alpha_1\int\lbrace{\nabla_\lambda}\left({{R_{\mu}}^{\nu\lambda\sigma}}(-)(\delta {\Gamma^\mu_{\nu\lambda}})\right)(\Theta(-n)) - ({{R_{\mu}}^{\nu\lambda\sigma}}(-))(\delta {\Gamma^\mu_{\nu\lambda}})(\delta(n){n_{,\lambda}})\\ - {\nabla_\lambda}({{R_{\mu}}^{\nu\lambda\sigma}}(-))(\delta {\Gamma^\mu_{\nu\lambda}})(\Theta(-n)) + ({{R_{\mu}}^{\nu\lambda\sigma}}(-))(\delta {\Gamma^\mu_{\nu\lambda}})(\delta(n){n_{,\lambda}})\rbrace\label{eq:120}
\end{multline}
And finally, we get:
\begin{align}
4\alpha_1\int\lbrace{\nabla_\lambda}\left({{R_{\mu}}^{\nu\lambda\sigma}}(-)(\delta {\Gamma^\mu_{\nu\lambda}})\right)(\Theta(-n)) - {\nabla_\lambda}({{R_{\mu}}^{\nu\lambda\sigma}}(-))(\delta {\Gamma^\mu_{\nu\lambda}})(\Theta(-n))\rbrace\label{eq:121}
\end{align}
We notice that taking a decomposition of the curvature (in the $n<0$ and $n>0$ regions) in the neighbourhood of our singular hypersurface $\Sigma_0$, the $\delta$-functions cancel out and do not manifest themselves, and this is correct since they were inherently absent from the beginning. The expression can further be rendered in terms of the discontinuity of the curvature across the  hypersurface:
\begin{align}
\left[{{R_{\mu}}^{\nu\lambda\sigma}}\right] = {{R_{\mu}}^{\nu\lambda\sigma}}(+) - {{R_{\mu}}^{\nu\lambda\sigma}}(-)\label{eq:122}
\end{align}
Using this, we can rephrase our integral \eqref{eq:115} as:
\begin{align}
-4\alpha_1\int_{\Sigma_0} \left[{{R_{\mu}}^{\nu\lambda\sigma}}\right] {\delta \Gamma^{\mu}_{\nu\sigma}}\,\sqrt{-g}\,d{S_{\lambda}} - 4\alpha_1\int\left({\nabla_\lambda}{{R_{\mu}}^{\nu\lambda\sigma}} + 2{A_{\lambda}}{{R_{\mu}}^{\nu\lambda\sigma}}\right){\delta \Gamma^{\mu}_{\nu\sigma}}\,\sqrt{-g}\,d^4 x\label{eq:123}
\end{align}
\subsubsection{\underline{[6.5.2.a]} Calculating discontinuities in various, necessary entities}
\label{subsubsec:jmps}
We will consider the case with discontinuities in the normal direction, because the Double Layer manifests itself through discontinuities in Curvature. We can accomplish this through changing the third index in \eqref{eq:123} and using $n$ the normal index instead:
\begin{align}
-4\alpha_1\int_{\Sigma_0} \left[{{R_{\mu}}^{\nu n\sigma}}\right] {\delta \Gamma^{\mu}_{\nu\sigma}}\,\sqrt{\lvert\gamma\rvert}\,d^3X \label{eq:124}
\end{align}
Negative sign appears due to the opposing directions of the normal whilst traversing the layer through the $-\hspace{1mm}(n<0)$ side to the $+\hspace{1mm}(n>0)$ side.\\
We put aside the bulk portion of this expression, to be consider at a later stage.
We now move on to calculate the necessary jumps, only the linear terms in the $\Gamma$ appear, not the second order terms $\Gamma\Gamma$ since their jumps are zero, refer to \eqref{eq:73} for details. In order to do this, we write out all the terms spanning the possible index permutations, respecting the symmetry with respect to the third index, so we cannot place two $n$ indices in the third and the fourth index, since that would result in a vanishing curvature tensor. These will be in Latin letters, since we are representing 3-dimensional jumps, as is evident from our implementation of the Gauss normal coordinates along with ADM (3+1) spacetime split \cite{MTW1973}. \textit{We will have three distinct kinds of jumps, four for the curvature tensor, four for the Ricci scalar and one for scalar curvature.} All of these shall be listed here and then calculated to be used later as needed.\\
$\bullet$ \textbf{For the curvature tensor}\\
With all of these considerations in place, we obtain four distinct jumps, for the curvature tensor, as given (We shall be adhering to  the completely contravariant version, with all indices upstairs):
\begin{align}
\left[{R^{nnni}}\right],\hspace{3mm}\left[{R^{ninj}}\right],\hspace{3mm}\left[{R^{linj}}\right],\hspace{3mm}\left[{R^{innj}}\right]
\label{eq:125}
\end{align}
$\bullet$ \textbf{For the Ricci tensor}\\
We have four different jumps:
\begin{align}
\left[{R^{nn}}\right],\hspace{3mm}\left[{R^{ni}}\right],\hspace{3mm}\left[{R^{in}}\right],\hspace{3mm}\left[{R^{ij}}\right]
\label{eq:126}
\end{align}
$\bullet$ \textbf{For the Scalar Curvature}\\
We have a single jump:
\begin{align}
\left[{R}\right]
\label{eq:127}
\end{align}
We now calculate all of these and list them all together so that we may use them later.
\newpage 
$\bullet$ \textbf{For the curvature tensor}:\\
$\circ$ $\left[{R^{nnni}}\right]$ (Using \cref{eq:94,eq:95}):
\begin{align}
&\left[{R^{n}}_{nni}\right] = \left[{{\Gamma^{n}}_{ni,n}}\right]\\ \nonumber
&{{\Gamma^{n}}_{ni}} = \cancelto{0}{{C^{n}}_{ni}} + {{W^{n}}_{ni}} = -\frac{1}{2}\left({A_{n}}{\delta^{n}_{i}} + {A_{i}}{\delta^{n}_{n}} - {A^{n}}{g_{ni}}\right) = -\frac{1}{2}(A_{i})\\ \nonumber
&\left[{{\Gamma^{n}}_{ni,n}}\right] = -\frac{1}{2}\left[{A_{i,n}}\right] + \frac{1}{2}\left[{A_{n,i}}\right] = -\frac{1}{2}\left[{F_{ni}}\right]\hspace{5mm}(\text{since}\,\left[{A_{n,i}}\right]\,\text{vanishes})\\ \nonumber
&\left[{R^{nnni}}\right] = \epsilon^2{g^{il}}\left\{-\frac{1}{2}\left[{F_{nl}}\right]\right\} = \epsilon^2\left\{\-\frac{1}{2}\left[{F_{n}}^{i}\right]\right\}\\ \nonumber
&\left[{R^{nnni}}\right] = -\frac{\epsilon}{2}\left[{F^{ni}}\right]
\label{eq:rnnni}
\end{align}
$\circ$ $\left[{R^{ninj}}\right]$:
\begin{align}
&\left[{R^{n}}_{inj}\right] = \left[{{\Gamma^{n}}_{ij,n}}\right]\\ \nonumber
&{{\Gamma^{n}}_{ij}} = {{C^{n}}_{ij}} + {{W^{n}}_{ij}} = \epsilon{K_{ij}} -\frac{1}{2}\left({A_{i}}{\delta^n_{j}} + {A_{j}}{\delta^n_{i}} - {A^{n}}{g_{ij}}\right)\\ \nonumber
&\left[{{\Gamma^{n}}_{ij,n}}\right] = \epsilon\left[{K_{ij,n}}\right] - \frac{1}{2}\left\{\left[{A_{i,n}}\right]\cancelto{0}{\delta^n_{j}} + \left[{A_{j,n}}\right]\cancelto{0}{\delta^n_{i}} - \epsilon\left[{A_{n,n}}\right]{g_{ij}}\right\}\\ \nonumber
&\left[{R^{ninj}}\right]= \left\{\epsilon^2{g^{il}}{g^{jp}}\left[{K_{lp,n}}\right] - \frac{1}{2}\left\{ - \epsilon^2\left[{A_{n,n}}\right]{g_{ij}}\right\}\right\}\\ \nonumber
&\left[{R^{ninj}}\right]= \left\{{g^{il}}{g^{jp}}\left[{K_{lp,n}}\right] + \frac{1}{2}\left\{\left[{A_{n,n}}\right]{g_{ij}}\right\}\right\}
\label{eq:rninj}
\end{align}
$\circ$ $\left[{R^{linj}}\right]$:
\begin{align}
&\left[{R^{linj}}\right] = \left[{\Gamma^{l}}_{ij,n}\right]\\ \nonumber
&{{\Gamma^{l}}_{ij}} = {{C^{l}}_{ij}} + {{W^{l}}_{ij}} = -\frac{1}{2}\left({A_{i}}{\delta^{l}}_{j} + {A_{j}}{\delta^{l}}_{i} - {A^{l}}{g_{ij}}\right)\\ \nonumber
&\left[{\Gamma^{l}}_{ij,n}\right] = -\frac{1}{2}\left(\left[{F_{ni}}\right]{{\delta^{l}}_{j}} + \left[{F_{nj}}\right]{\delta^{l}}_{i}- {g^{lm}}\left[{F_{nm}}\right]{g_{ij}}\right)\\ \nonumber
&\left[{R^{linj}}\right] = -\frac{1}{2}\left\{\left[{F^{ni}}\right]{g^{lj}} + \left[{F^{nj}}\right]{g^{li}}- \left[{F^{nl}}\right]{g^{ij}}\right\}
\label{eq:rlinj}
\end{align}
$\circ$ $\left[{R^{innj}}\right]$:
\begin{align}
\left[{R^{innj}}\right] = \left[{R^{ninj}}\right] = \left\{{g^{il}}{g^{jp}}\left[{K_{lp,n}}\right] + \frac{1}{2}\left\{\left[{A_{n,n}}\right]{g_{ij}}\right\}\right\}\\ \nonumber
\label{eq:rinnj}
\end{align}
Using the same calculation method, we evaluate the Ricci tensor jumps, using:\\
\begin{equation}
\left[{R_{\mu\nu}}\right] = \left[\frac{\partial {\Gamma^\lambda_{\mu\nu}}}{\partial {X^\lambda}}\right] - \left[\frac{\partial {\Gamma^\lambda_{\mu\lambda}}}{\partial {X^\nu}}\right]
\label{eq:ricjump}
\end{equation}
$\bullet$ \textbf{For the Ricci tensor}:\\
$\circ$ $\left[{R^{ij}}\right]$:
\begin{align}
\left[{R_{ij}}\right] = \left[\frac{\partial {\Gamma^n_{ij}}}{\partial n}\right] = \left[{\Gamma^n_{ij,n}}\right] = \epsilon\left({g^{il}}{g^{jp}}\left[K_{lp,n}\right] + \frac{1}{2}\left[A_{n,n}\right]{g^{ij}}\right) 
\label{eq:128}
\end{align}
$\circ$ $\left[{R^{in}}\right]$:
\begin{equation}
\left[{R^{in}}\right] = \frac{3}{2}\left[F^{ni}\right]
\label{eq:129}
\end{equation}
$\circ$ $\left[{R^{in}}\right]$:
\begin{equation}
\left[{R^{in}}\right] = -\frac{1}{2}\left[F^{ni}\right]
\label{eq:130}
\end{equation}
$\circ$ $\left[{R^{nn}}\right]$:
\begin{equation}
\left[{R^{nn}}\right] = \frac{1}{2}\left(2{g^{lp}}\left[{K_{lp,n}}\right] + 3\left[{A_{n,n}}\right]\right)
\label{eq:131}
\end{equation}
$\bullet$ \textbf{For the curvature scalar}:\\
\begin{equation}
\left[R\right] = {g_{ij}}\left[{R^{ij}}\right] = 2\left({g^{lp}}\left[{K_{lp,n}}\right] + \frac{3}{2}\left[{A_{n,n}}\right]\right)
\label{eq:132}
\end{equation}
\subsubsection{\underline{[6.5.2.b]} A quick explanatory detour}
\label{subsubsec:detour}
We take a quick detour to elucidate the steps which are to follow. For the intents and purposes of this thesis, we shall be limiting ourselves to a few noteworthy variations, there are a few reasons for choosing to do this. Since we are interested in novel effects which aid us in distinguishing how Weyl Gravity has physical and mathematical structure in place to help us sustain our search for Double Layers and associated physical effects. We also wish to learn more specifically about the nature of surface energy momentum effects (\ref{sec:vol}) connected to our hypersurface and relations between the source field present on the surface and the Weyl one-form which gauges Weyl gravity. All of this is not plausible in GR. If we find such a relation to exist (which we shall talk about later in \ref{chap:conclude}), we can establish a distinctive difference between the behaviour of GR and Weyl gravity, besides the obvious mathematical structure in place. The variations we shall be focussing on are: $\delta {A_i}, \delta {A_n}, \delta {g_{nn}}, (\delta {g_{ij}})_{,n}, \delta {g_{ni}}$.\\
You might pose the question as to why we don't investigate the $\delta{g_{ij}}$ variation, since we consider its partial derivative with respect to the normal. There is a specified impetus behind this which hinges the entire crux of this thesis and is of profound significance.
\subsubsection{\underline{[6.5.2.c]} Reasons for not considering the $\delta{g_{ij}}$ variation}
\begin{enumerate}[label =\roman*.]
 \item $\delta {g_{ij}} = \delta {\gamma_{ij}}$, due to our choice of Gauss normal coordinate system \ref{sec:gncs}.
 \item $\lbrace\cdots\rbrace\left(\delta {\gamma_{ij,n}}\right) = -2\lbrace\cdots\rbrace\left(\delta {K_{ij}}\right)$, the variation in the extrinsic curvature can be expressed in terms of the surface metric. So using $\delta {g_{ij}}$ is superfluous.
 \item $\delta {K_{ij}} = {{B_{ij}}^{{i}^{\prime}{j}^{\prime}}}{\delta\gamma_{{i}^{\prime}{j}^{\prime}}}$, Here, ${{B_{ij}}^{{i}^{\prime}{j}^{\prime}}}$- is a selectively arbitrary depending only on the choice of the solutions in the $(\pm)$ regions.
 \item The equations: 
        \begin{align*}
         \lbrace\cdots\rbrace\delta {g_{ij}} = {{S}^{ij}}{\delta {g_{ij}}}\,(+\frac{1}{2})
        \end{align*}
        serve to define ${{B_{ij}}^{{i}^{\prime}{j}^{\prime}}}$. They give us no information about the structure of the double layer itself, i.e. information about the jumps: $\left[{K_{lp,n}}\right]$, $\left[{A_{n,n}}\right]$, $\left[{F^{ni}}\right]$.
 \item ${{B_{ij}}^{{i}^{\prime}{j}^{\prime}}}$ does not enter the equations with $\delta {A_{i}}$, $\delta {A_{n}}$, $\delta {g_{nn}}$, $\delta {g_{ni}}$
\end{enumerate}
Considering all of these, we can safely neglect $\delta {g_{ij}}$ in our calculations.
\subsubsection{\underline{[6.5.2.d]} Collecting the Double Layer contributing variation terms for $\alpha_1$}
We shall be considering the variation of the following quantities:
\begin{center}
$\delta {A_{n}}$\hspace{3mm} $\delta {A_{i}}$\hspace{3mm} $(\delta {g_{ij}})_{,n}$\hspace{3mm} $(\delta {g_{nn}})$\hspace{3mm} $(\delta {g_{ni}})$
\end{center}
Since $\Gamma^{\mu}_{\nu\sigma}$ can be decomposed into its Riemannian connection (Christoffel symbols) and Weylian connection parts:
\begin{align*}
{\Gamma^{\mu}}_{\nu\sigma} &= {C^{\mu}}_{\nu\sigma} + {W^{\mu}}_{\nu\sigma}\\
{\delta}{\Gamma^{\mu}}_{\nu\sigma} &= {\delta}{C^{\mu}}_{\nu\sigma} + {\delta}{W^{\mu}}_{\nu\sigma}
\end{align*}
Ergo, we can considerably simplify the variation by breaking the connection up as shown above and in \eqref{eq:76}. We first consider the Weylian portion since it is significantly greater in difficulty to calculate and later take the metric case.
\subsubsection{\underline{[6.5.2.e]} The Weylian portion}
We consider the integrand in \eqref{eq:124}:
\begin{align}
\left[{R_{\mu}}^{\nu n \sigma}\right]{(\delta {W^{\mu}_{\nu\sigma}})} \label{eq:134}
\end{align}
Using \cref{eq:95}, we can write down the variation for the Weyl connection and take relevant products to get our variations, we write down the generalised 4-dimensional version of the integrand and then implement the (3+1) split, considering all possible permutations with the normal coordinate occupying various permissible indices:
\begin{dmath}
-\frac{1}{2}\left\{\left[{{R_{\mu}}^{\nu n \sigma}}\right]\left\{(\delta{A_{\nu}}){\delta^{\mu}_{\sigma}} + (\delta{A_{\sigma}}){\delta^{\mu}_{\nu}} - {(\delta {A_{\kappa}})}{g^{\kappa\mu}}{g_{\nu\sigma}} + {A_{\kappa}}{g_{\nu\sigma}}{g^{\mu\alpha}}{g^{\kappa\rho}}(\delta {g_{\alpha\rho}}) - {A_{\kappa}}{g^{\mu\kappa}}(\delta {g_{\nu\sigma}})\right\}\right\}\\ \nonumber
=-\frac{1}{2}\left\{\left[{{R_{\sigma}}^{\nu n \sigma}}\right](\delta {A_{\nu}}) + \left[{{R_{\nu}}^{\nu n \sigma}}\right](\delta {A_{\sigma}}) - \left[{R^{\kappa\nu n\sigma}}\right]{g_{\nu\sigma}}{(\delta{A_{\kappa}})} + \left[{R^{\alpha\nu n\sigma}}\right]{g_{\nu\sigma}}{A_{\kappa}}{g^{\kappa\rho}}(\delta g_{\alpha\rho}) - \left[{R^{\kappa\nu n \sigma}}\right]{A_{\kappa}}(\delta g_{\nu\sigma})\right\}
\label{eq:135}
\end{dmath}
We now transition the relation \cref{eq:135} to our (3+1) ADM split along with the prescribed gauss normal coordinate system (see Section \ref{sec:gncs}) whilst also being privy to the fact that according to \cref{eq:90} the terms with $g_{ni}$ and $g^{ni}$ vanish and $g_{nn} = g^{nn} = \epsilon = \pm 1$:
\begin{dmath}
=-\frac{1}{2}\left\{\left[{R^{linj}}\right]{g_{lj}}(\delta {A_{i}}) + \left[{R^{linj}}\right]{g_{li}}(\delta {A_{j}}) - \left[{R^{linj}}\right]{g_{ij}}{(\delta{A_{l}})} + \left[{R^{ilnm}}\right]{g_{lm}}{A_{p}}{g^{pj}}(\delta g_{ij}) - \left[{R^{linj}}\right]{A_{l}}(\delta g_{ij}) - \left[{R^{ninj}}\right]{A_{n}}(\delta g_{ij})\right\}\\ 
-\frac{1}{2}\left\{\left[{R^{lnnj}}\right]{g_{lj}}(\delta {A_{n}}) + \left[{R^{lnnj}}\right]{g_{li}}(\delta {A_{n}}) - \left[{R^{lnni}}\right]{A_{l}}{(\delta{g_{ni}})} + \left[R^{nlnp}\right]{g_{lp}}{A_{j}}{g^{ji}}\left(\delta g_{ni}\right) +\epsilon\left[R^{ninj}\right]{g_{ij}}{A_{n}}\left(\delta g_{nn}\right) - \left[R^{innn}\right]{A_{i}}\left(\delta g_{nn}\right)\right\}
\label{eq:136}
\end{dmath}
Further, we take permutations with the normal index in other places besides the middle index in the curvature tensor, along with correct contractions with the Weylian variation terms:
\begin{align}
&=-\frac{1}{2}\left\{\left[{{R_{n}}^{\nu n \sigma}}\right]\left\{(\delta{A_{\nu}}){\delta^{n}_{\sigma}} + (\delta{A_{\sigma}}){\delta^{n}_{\nu}} - {(\delta {A_{\kappa}})}{g^{\kappa n}}{g_{\nu\sigma}} + {A_{\kappa}}{g_{\nu\sigma}}{g^{n\alpha}}{g^{\kappa\rho}}(\delta {g_{\alpha\rho}}) - {A_{\kappa}}{g^{n\kappa}}(\delta {g_{\nu\sigma}})\right\}\right\}\\ \nonumber
&=-\frac{1}{2}\left\{\left[{{R_{n}}^{\nu n \sigma}}\right]\left\{(\delta{A_{\nu}}){\delta^{n}_{\sigma}} + (\delta{A_{\sigma}}){\delta^{n}_{\nu}}\right\}\right\}\\ \nonumber
\label{eq:137}
\end{align}
Again, we use the (3+1) ADM split, as we did before, to get:
\begin{align}
&= -\frac{\epsilon}{2}\left\{\left[R^{ninj}\right]\left\{(\delta A_{i})\cancelto{0}{\delta^{n}_{j}} + (\delta A_{j})\cancelto{0}{\delta^{n}_{i}}\hspace{3mm}\right\}\right\} -\frac{\epsilon}{2}\left\{\left[R^{nnni}\right]\left\{(\delta A_{n})\cancelto{0}{\delta^{n}_{i}} + (\delta A_{i}){\delta^{n}_{n}}\right\}\right\}\\ \nonumber
&= -\frac{\epsilon}{2}\left\{\left[R^{nnni}\right]\left\{(\delta A_{i})\right\}\right\}\\
\label{eq:138}
\end{align}

Now let us collect the coefficients for the variations, one by one:\medskip

\hspace{0.2mm}$\circ\hspace{3mm}\delta {A_{i}}$:
\begin{align}
\frac{3}{4}\left\{\left[{F^{ni}}\right]\right\}
\label{eq:139}
\end{align}

$\circ\hspace{3mm}\delta {A_{n}}$: 
\begin{align}
2\left\{{g^{pl}}\left[{K_{lp,n}}\right] + \frac{3}{2}\left[{A_{n,n}}\right]\right\}
\label{eq:140}
\end{align}

$\circ\hspace{3mm}\delta {g_{ij}}$: 
\begin{align}
\frac{1}{2}\left\{\left[{F^{nl}}_{|l}\right]{g^{ij}} - 2{K^{jl}}\left[{F^{ni}}_{|l}\right]\right\} + \left\{{g^{il}}{g^{jp}}\left[{K_{lp,n}}\right] + \frac{1}{2}\left[{A_{n,n}}\right]{g^{ij}}\right\}{A_{n}}
\label{eq:141}
\end{align}

$\circ\hspace{3mm}\delta {g_{ni}}$:
\begin{align}
\frac{1}{2}\left\{{g^{il}}{g^{jp}}\left[{K_{lp,n|j}}\right] + \frac{1}{2}\left[{A_{n,n|j}}\right]{g^{ij}}\right\}
\label{eq:143}
\end{align}

$\circ\hspace{3mm}\delta {g_{nn}}$:
\begin{align}
-\frac{\epsilon}{2}\left\{{g^{lp}}\left[{K_{lp,n}}\right] + \frac{3}{2}{\left[A_{n,n}\right]}\right\}{A_{n}}
\label{eq:144}
\end{align}

Here we have gathered the 3-dimensional variation coefficients, adding these to those obtained from the Christoffel portion and the contributions to the surface from the bulk, we can ascertain the final coefficients.

\subsubsection{\underline{[3.3.2.f]} The Christoffel portion}
As we did in the Weylian case, we collect the coefficients of the relevant variations and proceed henceforth much as we did above:
\begin{align}
\left[{R_{\mu}}^{\nu n \sigma}\right]{(\delta {{C^{\mu}}_{\nu\sigma}})} \label{eq:145}
\end{align}
Like before, we take this and expand using \cref{eq:94}. This is calculated as follows:\newline
\begin{align}
&\left\{\left[{R_{\mu}}^{\nu n \sigma}\right]{(\delta {{C^{\mu}}_{\nu\sigma}})}\right\} = \frac{1}{2}\left\{\left[{R_{\mu}}^{\nu n \sigma}\right]{g^{\mu\kappa}}\left( (\delta g_{\kappa\nu})_{;\sigma} + (\delta {g_{\kappa\sigma}})_{;\nu} - (\delta {g_{\nu\sigma}})_{;\kappa}\right)\right\}\\
&= \frac{1}{2}\left\{\left[R^{\kappa\nu n \sigma}\right]\left( (\delta g_{\kappa\nu})_{;\sigma} + (\delta {g_{\kappa\sigma}})_{;\nu} - (\delta {g_{\nu\sigma}})_{;\kappa}\right)\right\}
\end{align}
The sum of the covariant derivatives of the variation terms of the metric can be simplified as such:
\begin{align}
&+(\delta {g_{\kappa\nu}})_{;\sigma} = (\delta {g_{\kappa\nu}})_{,\sigma}  + {C^{\rho}_{\sigma\kappa}}(\delta {g_{\nu\rho}}) + {C^{\rho}_{\sigma\nu}}(\delta {g_{\kappa\rho}})\\ \nonumber
&+(\delta {g_{\kappa\sigma}})_{;\nu} = (\delta {g_{\kappa\sigma}})_{,\nu} + {C^{\rho}_{\nu\kappa}}(\delta {g_{\sigma\rho}}) + {C^{\rho}_{\nu\sigma}}(\delta {g_{\kappa\rho}})\\ \nonumber
&-(\delta {g_{\nu\sigma}})_{;\kappa} = (\delta {g_{\nu\sigma}})_{,\kappa} + {C^{\rho}_{\kappa\nu}}(\delta {g_{\sigma\rho}}) + {C^{\rho}_{\kappa\sigma}}(\delta {g_{\nu\rho}})\\ \nonumber
\implies &((\delta g_{\kappa\nu})_{;\sigma} + (\delta {g_{\kappa\sigma}})_{;\nu} - (\delta {g_{\nu\sigma}})_{;\kappa}) = (\delta {g_{\kappa\nu}})_{,\sigma} + (\delta {g_{\kappa\sigma}})_{,\nu} - (\delta {g_{\nu\sigma}})_{,\kappa} + 2{C^{\rho}_{\nu\sigma}}(\delta {g_{\kappa\rho}})
\end{align}
Consequently, our integrand takes the form:
\begin{align}
\frac{1}{2}\left\{\left[R^{\kappa\nu n \sigma}\right]\left( (\delta {g_{\kappa\nu}})_{,\sigma} + (\delta {g_{\kappa\sigma}})_{,\nu} - (\delta {g_{\nu\sigma}})_{,\kappa} + 2{C^{\rho}_{\nu\sigma}}(\delta {g_{\kappa\rho}})\right)\right\}
\end{align}
Using the above relation, we can get the (3+1) form for it:
\begin{dmath}
-\frac{1}{2}\left[R^{ninj}\right](\delta {g_{ij}})_{,n} + \left[R^{innl}\right](C^{j}_{nl})(\delta {g_{ij}}) + \left[R^{imnl}\right](C^{j}_{ml})(\delta {g_{ij}}) + \left[R^{ninj}\right](C^{\,n}_{ij})(\delta {g_{nn}}) + \left[R^{nlnj}\right](C^{\,i}_{lj})(\delta {g_{ni}}) + \left[{R^{nnnl}}\right](C^{\,i}_{nl})(\delta {g_{ni}}) - \left[{R^{linj}}\right](C^{\,n}_{lj})(\delta {g_{ni}})
\end{dmath}
Using this, we can write down our variation coefficients for the Christoffel portion:\newline
$\circ\hspace{3mm}(\delta {g_{ij}})_{,n}$:
\begin{align}
-\frac{1}{2}\left\{{g^{il}}{g^{jp}}\left[{K_{lp,n}}\right] + \frac{1}{2}\left[{A_{n,n}}\right]{g^{ij}}\right\}
\label{eq:146}
\end{align}
$\circ\hspace{3mm}\delta {K_{ij}}$:
\begin{align}
\frac{1}{2}\left\{{g^{il}}{g^{jp}}\left[{K_{lp,n}}\right] + \frac{1}{2}\left[{A_{n,n}}\right]{g^{ij}}\right\}
\label{eq:al1chriKij}
\end{align}
$\circ\hspace{3mm}\delta {g_{ij}}$: 
\begin{dmath}
-\frac{1}{2}\left\{\left[{F^{ni}}\right]{g^{lj}} + \left[{F^{nj}}\right]{g^{li}} - \left[{F^{nl}}\right]{g^{ij}} \right\} -\\ \frac{1}{2}\left\{ \left[{F^{nj}}_{|m}\right]{g^{im}} + \left[{F^{nj}}_{|l}\right]{g^{li}} - \left[{F^{ni}}_{|m}\right]{g^{jm}} - \left[{F^{ni}}_{|l}\right]{g^{jl}}\right\} +\\ \left\{{K^{lp}}\left[{K_{lp,n}}\right] + \frac{1}{2}\left[{A_{n,n}}\right]{K^{ij}}\right\}
\label{eq:al1chrisgij}
\end{dmath}
$\circ\hspace{3mm}\delta {g_{nn}}$:
\begin{align}
\left\{{K^{lp}}\left[{K_{lp,n}}\right] + \frac{1}{2}{K^{lp}}{g_{lp}}\left[{A_{n,n}}\right]\right\}
\label{eq:147}
\end{align}
$\circ\hspace{3mm}\delta {g_{ni}}$:
\begin{align}
\left\{{K^{lp}}\left[{K_{lp,n|i}}\right] + \frac{1}{2}{K^{lp}}{g_{lp}}\left[{A_{n,n|i}}\right]\right\} -\frac{\epsilon}{2}{K^{i}_{l}}\left[{F^{nl}}\right] +\frac{1}{2}\left\{K{\left[{F^{ni}}\right]} +{K^{l}_{j}}{\left[{F^{nj}}\right]} - {K^{i}_{l}}{\left[{F^{nl}}\right]}\right\}
\label{eq:148}
\end{align}
As can be seen, no ($\delta A$) terms appear in the Christoffel coefficient part of the variation process, as is inevitable from the definition of the variation of the Christoffel symbols \cref{eq:94}, hence their absence is justified. Now we consider contributions from the bulk. 

\subsubsection{\underline{[3.3.2.g]} Bulk contribution portion of the \texorpdfstring{$\alpha_1$}{alpha 1} variation}
So far we have only considered one of the two integrals associated with the $\alpha_1$ term variation, namely, \cref{eq:125}, we also need to consider the bulk portion of this term, (i.e. the second integral which we set aside for later), before we can collect all associated terms of the total variation for $\alpha_1$. this integral as we know from \cref{eq:124}, is:
\begin{equation}
- 4\alpha_1\int \left({\nabla_\lambda}{{R_{\mu}}^{\nu\lambda\sigma}} + 2{A_{\lambda}}{{R_{\mu}}^{\nu\lambda\sigma}}\right){\delta \Gamma^{\mu}_{\nu\sigma}}\,\sqrt{-g}\,d^4 x
\label{eq:149}
\end{equation}
Since, $\left({\nabla_\lambda}{{R_{\mu}}^{\nu\lambda\sigma}} + 2{A_{\lambda}}{{R_{\mu}}^{\nu\lambda\sigma}}\right)$, has a contraction only across the $\lambda$ index, we can rewrite this derivative as a newly defined tensor entity: 
\begin{align}
{D_{\mu}}^{\nu\sigma} = \left({\nabla_\lambda}{{R_{\mu}}^{\nu\lambda\sigma}} + 2{A_{\lambda}}{{R_{\mu}}^{\nu\lambda\sigma}}\right) 
\label{eq:D}
\end{align}
yielding us:
\begin{align}
- 4\alpha_1\int\left({D_{\mu}}^{\nu\sigma}\right)({\delta \Gamma^{\mu}_{\nu\sigma}})\,\sqrt{-g}\,d^4x
\label{eq:150}
\end{align}
Again, as we did in the previous two Subsubsections, we invoke the relation \cref{eq:76}, to give us:
\begin{align}
&- 4\alpha_1\int\left({D_{\mu}}^{\nu\sigma}\right)({\delta W^{\mu}_{\nu\sigma}} + {\delta C^{\mu}_{\nu\sigma}})\,\sqrt{-g}\,d^4x \\ \nonumber
&= -4\alpha_1\int\left({D_{\mu}}^{\nu\sigma}\right)({\delta W^{\mu}_{\nu\sigma}})\,\sqrt{-g}\,d^4x - 4\alpha_1\int\left({D_{\mu}}^{\nu\sigma}\right)({\delta C^{\mu}_{\nu\sigma}})\,\sqrt{-g}\,d^4x
\end{align}
Albeit, for our intents and purposes, we shall not consider the Weylian part of this equation, since it contributes only to the field equations of the Bulk, hence, we have:
\begin{align}
- 4\alpha_1\int\left({D_{\mu}}^{\nu\sigma}\right)({\delta C^{\mu}_{\nu\sigma}})\,\sqrt{-g}\,d^4x
\end{align}
We know from \cref{eq:D} and \cref{eq:91}, that the integrand in the above equation can be re-written as:
\begin{align}
\left({D_{\mu}}^{\nu\sigma}\right)({\delta C^{\mu}_{\nu\sigma}}) &= \left({D_{\mu}}^{\nu\sigma}\right)\left\{\frac{1}{2}{(g^{\mu\kappa})}\left( (\delta {g_{\kappa\nu}})_{;\sigma} + (\delta {g_{\kappa\sigma}})_{;\nu} -(\delta {g_{\nu\sigma}})_{;\kappa}\right)\right\}\\ \nonumber
&= \frac{1}{2}\left\{\left( {D^{\kappa\nu\sigma}}(\delta {g_{\kappa\nu}})_{;\sigma} + {D^{\kappa\nu\sigma}}(\delta {g_{\kappa\sigma}})_{;\nu} -{D^{\kappa\nu\sigma}}(\delta {g_{\nu\sigma}})_{;\kappa}\right)\right\}
\end{align}
As implemented in the previous Subsubsections, we again perform the (3+1) split, to get:
\begin{align}
&=\frac{1}{2}\left\{\left( {D^{nin}}(\delta {g_{ni}})_{;n} + {D^{nin}}(\delta {g_{nn}})_{;i} -{D^{nin}}(\delta {g_{in}})_{;n}\right)\right\}\\ \nonumber
&+\frac{1}{2}\left\{\left( {D^{inn}}(\delta {g_{in}})_{;n} + {D^{inn}}(\delta {g_{in}})_{;n} -{D^{inn}}(\delta {g_{nn}})_{;i}\right)\right\} + \frac{1}{2}\left\{(\delta {g_{nn}})_{;n}\right\}\\ \nonumber
&= \frac{1}{2}\left\{(\delta {g_{nn}})_{;n}\right\}{D^{nnn}} + \frac{1}{2}\left\{2(\delta {g_{ni}})_{;n}\right\}{D^{inn}} \\ \nonumber
&= \frac{1}{2}\left\{(\delta {g_{nn}})_{;n}{D^{nnn}} + 2(\delta {g_{ni}})_{;n}{D^{inn}}\right\}
\end{align}
We know that we can rewrite this as a difference between two derivatives as so:
\begin{align}
\frac{1}{2}\left\{ ({D^{nnn}}(\delta {g_{nn}}))_{;n} + 2({D^{inn}}(\delta {g_{ni}}))_{;n}\right\} - \frac{1}{2}\left\{(\delta {g_{nn}}){D^{nnn}}_{;n} + 2(\delta {g_{ni}}){D^{inn}}_{;n}\right\}
\end{align}
Consequently, we get the discontinuities as given:
\begin{align}
\frac{1}{2}\left\{ \left[{D^{nnn}}\right](\delta {g_{nn}}) + 2\left[{D^{inn}}\right](\delta {g_{ni}})\right\}
\end{align}
We can modify \cref{eq:D} to get a version with an entirely contravariant $\left[{D^{\mu\nu\sigma}}\right]$:
\begin{align}
{D_{\mu}}^{\nu\sigma} &= {\nabla_\lambda}{{R_{\mu}}^{\nu\lambda\sigma}} + 2{A_{\lambda}}{{R_{\mu}}^{\nu\lambda\sigma}}\\ \nonumber
&= {\nabla_\lambda}({{g_{\mu\kappa}}R^{\kappa\nu\lambda\sigma}}) + 2{A_{\lambda}}({{g_{\mu\kappa}}R^{\kappa\nu\lambda\sigma}})\\ \nonumber
&= {g_{\mu\kappa}}\left\{{\nabla_\lambda}({R^{\kappa\nu\lambda\sigma}}) + 3{A_{\lambda}}({R^{\kappa\nu\lambda\sigma}})\right\}\\ \nonumber
&{D}^{\mu\nu\sigma} = \left\{{\nabla_\lambda}({R^{\mu\nu\lambda\sigma}}) + 3{A_{\lambda}}({R^{\mu\nu\lambda\sigma}})\right\}
\end{align}

We can now evaluate the discontinuities $\left[D^{nnn}\right]$ and $\left[D^{inn}\right]$ using, to get the $\delta {g_{nn}}$ and $\delta {g_{ni}}$ variation terms:\newline

$\bullet\hspace{2mm}\left[D^{nnn}\right]$ (the coefficient for $\delta {g_{ni}}$):
\begin{align}
\left[D^{nnn}\right] &= {\nabla_{l}}{{R}^{nnln}} + 3{A_{l}}{{R}^{nnln}}\\ \nonumber
&= -\frac{\epsilon}{2}\left\{ [{F^{nl}}_{|l}] + {A_{l}}[{F^{nl}}]\right\}
\end{align}
$\bullet\hspace{2mm}\left[D^{inn}\right]$ (the coefficient for $\delta {g_{nn}}$):
\begin{dmath}
\left[D^{inn}\right] = \left\{{\nabla_{l}}{{R}^{inln}} + 3{A_{l}}{{R}^{inln}}\right\} + \left\{{\nabla_{n}}{{R}^{innn}} + 3{A_{n}}{{R}^{innn}}\right\}\\ 
= \left\{ {g^{ij}}{g^{lp}}\left[{K_{jp,n|l}}\right] + \frac{1}{2}{g^{il}}\left[{A_{n,n|l}}\right] \right\} + \frac{A^{i}}{4}\left\{2{g^{jp}}\left[{K_{jp,n}}\right] + 3\left[{A_{n,n}}\right]\right\} + \frac{\epsilon}{2}\left\{{K^{l}_{i}}\left[{F^{ni}}\right] - (K + {A_{n}})\left[{F^{nl}}\right]\right\}
\end{dmath}

\subsubsection{\underline{[6.5.2.h]} Final terms of the $\alpha_1$ variation}
Now we can use these four dimensional jumps, implement our Gauss normal coordinate system and check for all three dimensional symmetries around the third index which has to be $n$, by doing this, we attain the following coefficients for our variations (these involve using the jumps we evaluated in \ref{subsubsec:jmps} and compiling different results from the variation coefficients above so as to obtain the correct final variation coefficients for our case), these involve protracted calculations and I shall be introducing the results we obtained directly due to page limit constraints and for the greater brevity:\newline
$\rightarrow\hspace{3mm}\delta {A_{n}}$:
\begin{align}
\left(2{g^{pl}}\left[{K_{lp,n}}\right] + 3\left[{A_{n,n}}\right]\right)(\delta{A_{n}})
\label{eq:al1an}
\end{align}
$\rightarrow\hspace{3mm}\delta {A_{i}}$:
\begin{align}
 \frac{3}{4}\left[{F^{ni}}\right](\delta{A_{i}})
\label{eq:al1ai}
\end{align}
$\rightarrow\hspace{3mm}(\delta {g_{ij}})_{,n}$:
\begin{align}
-\frac{1}{2}\left(2{g^{il}}{g^{jp}}\left[{K_{lp,n}}\right] + {g^{ij}}\left[{A_{n,n}}\right]\right)(\delta {g_{ij}})_{,n}
\label{eq:al1gijn}
\end{align}
$\rightarrow\hspace{3mm}\delta {g_{nn}}$:
\begin{dmath}
\left\{ -\frac{\epsilon}{2}\left(\frac{1}{2}\left[{F^{nl}}_{|l}\right] + 2{K^{lp}}\left[{K_{lp,n}}\right] + K\left[{A_{n,n}}\right]\right) -\frac{\epsilon}{4}{A_{n}}\left(2{g^{pl}}\left[{K_{lp,n}}\right] + 3\left[{A_{n,n}}\right]\right) + \frac{\epsilon}{2}\left(\frac{1}{2}\left[{F^{nl}}_{|l}\right] + {A_{l}}\left[{F^{nl}}\right]\right)\right\}(\delta {g_{nn}})
\label{eq:al1gnn}
\end{dmath}
$\rightarrow\hspace{3mm}\delta {g_{ni}}$:
\begin{dmath}
\left\{\frac{3}{2}\left(2{g^{il}}{g^{jp}}\left[{K_{lp,n|j}}\right] + {g^{ij}}\left[{A_{n,n|j}}\right]\right) +{A^{i}}\left(2{g^{pl}}\left[{K_{lp,n}}\right] + 3\left[{A_{n,n}}\right]\right)\\ + {\epsilon}\left(-\frac{5}{2}{{K^i}_l}{\left[{F^{nl}}\right]} + K{\left[{F^{ni}}\right]}\right)\right\}(\delta{g_{ni}})
\label{eq:al1gni}
\end{dmath}
I have cut short certain similar calculations in order to make this thesis self-consistent, such that all the four $\alpha_{i},\hspace{0.85mm} i\in \lbrace1,2,3,4\rbrace$ may be discussed so that the reader may have a concise understanding of the entire cohesive picture and how the variations fit into our investigations into Double Layer physics, without exceeding the bounds of the page limit constraint. Taking all this into account, I shall not get into extensive detail as to how we get the $\alpha_2$ and $\alpha_3$ variation coefficients, rather I shall just list the results which I obtained in a quest to calculate them, following this, we put all of them together in one single integral, which yields us the terms relevant to the Double Layer in the Weyl Action, we can then compare them with the surface and charge action on the R.H.S in a special case, considering a vacuum and draw important conclusions therein. Onto the $\alpha_2$ and $\alpha_3$ cases which are now briefly discussed.
\subsection{Variation of the \texorpdfstring{$\alpha_2$}{alpha 2} term}
\label{subsec:al2}
Commencing the original integral as given:
\begin{align}
2\alpha_2\int {R^{\mu\nu}}{\delta R_{\mu\nu}}\,\sqrt{-g}\,d^4X
\label{eq:alpha2}
\end{align}
 
The $\alpha_2$ variation is very similar in its premise to the way we had setup the $\alpha_1$ variation, with the use of Palatini's Formalism;
\begin{align}
2\alpha_2\int {R^{\mu\nu}}\left({{\nabla_\lambda}(\delta {{\Gamma^\lambda}_{\mu\nu}}) - {\nabla_\nu}(\delta {{\Gamma^\lambda}_{\mu\lambda}})}\right)\,\sqrt{-g}\,d^4X
\label{eq:152}
\end{align}
Performing the first iteration of the Stokes theorem and following the directions in \ref{sec:plan}, we will get:
\begin{align}
2\alpha_2\int_{\Sigma_0}\left[{R^{\mu\nu}}\right]\left({{\nabla_\lambda}(\delta {{\Gamma^\lambda}_{\mu\nu}}) - {\nabla_\nu}(\delta {{\Gamma^\lambda}_{\mu\lambda}})}\right)\,\sqrt{\lvert\gamma\rvert}\,d^3X
\label{eq:153}
\end{align}
The discontinuities we saw in \ref{subsubsec:jmps}, appear here, in the final form of this calculation, albeit I shall replicate here the final results of the calculations:\newline
$\rightarrow\hspace{3mm}\delta {A_{n}}$:
\begin{align}
\left(2{g^{pl}}\left[{K_{lp,n}}\right] + 3\left[{A_{n,n}}\right]\right)(\delta{A_{n}})
\label{eq:al2an}
\end{align}
$\rightarrow\hspace{3mm}\delta {A_{i}}$:
\begin{align}
 \frac{5}{2}\left[{F^{ni}}\right](\delta{A_{i}})
\label{eq:al2ai}
\end{align}
$\rightarrow\hspace{3mm}(\delta {g_{ij}})_{,n}$:
\begin{align}
-\frac{1}{2}\left(({g^{ij}}{g^{lp}} + {g^{il}}{g^{jp}})\left[{K_{lp,n}}\right] + 2{g^{ij}}\left[{A_{n,n}}\right]\right)(\delta {g_{ij}})_{,n}
\label{eq:al2gijn}
\end{align}
$\rightarrow\hspace{3mm}\delta {g_{nn}}$:
\begin{dmath}
-\frac{\epsilon}{4}\left(\left[{F^{nl}}_{|l}\right] - \frac{3}{2}{A_{l}}\left[{F^{nl}}\right] + (2K + 3{A_n})\left(2{g^{lp}}\left[{K_{lp,n}}\right] + 3\left[{A_{n,n}}\right]\right)\right) (\delta {g_{nn}})
\label{eq:al2gnn}
\end{dmath}
$\rightarrow\hspace{3mm}\delta {g_{ni}}$:
\begin{dmath}
\frac{1}{2}\left\{{\epsilon}\left(-2{{K^i}_l}{\left[{F^{nl}}\right]} + 3{A_n}{\left[{F^{ni}}\right]}\right) -\left(2{g^{il}}{g^{jp}}\left[{K_{lp,n|j}}\right] + {g^{ij}}\left[{A_{n,n|j}}\right]\right)\\+{A^{i}}\left(2{g^{pl}}\left[{K_{lp,n}}\right] + 3\left[{A_{n,n}}\right]\right) \right\}(\delta{g_{ni}})
\label{eq:al2gni}
\end{dmath}
With these at hand, we finally face our final and quite considerably one of the least complicated variations.
\subsection{Variation of the \texorpdfstring{$\alpha_3$}{alpha 3} term}
\label{subsec:al3}
Once again, we start with the original integral as given:
\begin{align}
2\alpha_3\int {R}{g^{\mu\nu}}{\delta R_{\mu\nu}}\,\sqrt{-g}\,d^4X
\label{eq:154}
\end{align}
And sticking to the strategy we outlined in \ref{sec:plan}, we arrive at:
\begin{align}
2\alpha_3\int {R}{g^{\mu\nu}}\left({{\nabla_\lambda}(\delta {{\Gamma^\lambda}_{\mu\nu}}) - {\nabla_\nu}(\delta {{\Gamma^\lambda}_{\mu\lambda}})}\right)\,\sqrt{-g}\,d^4X
\label{eq:155}
\end{align}
Forming the total derivative with respect to the covariant derivative:
\begin{dmath}
- 2\alpha_3\int \left\{\left\{\left({R}{g^{\mu\nu}}(\delta {{\Gamma^\lambda}_{\mu\nu}})\right)_{;\lambda} + {2A_\lambda}\left({R}{g^{\mu\nu}}(\delta {{\Gamma^\lambda}_{\mu\nu}})\right)\right\} - \left\{\left({R}{g^{\mu\nu}}(\delta {{\Gamma^\lambda}_{\mu\lambda}})\right)_{;\nu}+ {2A_\nu}\left({R}{g^{\mu\nu}}(\delta {{\Gamma^\lambda}_{\mu\lambda}})\right)\right\}\right\}\,\sqrt{-g}\,d^4X
- 2\alpha_3\int \left\{{\nabla_\lambda}({R}{g^{\mu\nu}}){\delta {{\Gamma^\lambda}_{\mu\nu}}} - {\nabla_\nu}({R}{g^{\mu\nu}}){\delta {{\Gamma^\lambda}_{\mu\lambda}}}\right\}\,\sqrt{-g}\,d^4X
\label{eq:156}
\end{dmath}
Further, to simplify the second integral, we have:
\begin{dmath}
{\nabla_\lambda}({R}{g^{\mu\nu}}) = {g^{\mu\nu}}(\nabla_\lambda R) - {A_\lambda}{R}{g^{\mu\nu}}
{\nabla_\lambda}({R}{g^{\mu\nu}}) + 2{A_\lambda}{R}{g^{\mu\nu}} = {g^{\mu\nu}}(\nabla_\lambda R) + {A_\lambda}{R}{g^{\mu\nu}}
= (\nabla_\lambda R + {A_\lambda}{R}){g^{\mu\nu}}
= {H_\lambda}{g^{\mu\nu}} \label{eq:157}
\end{dmath}
Here we defined a new tensor entity to make dealing with the integral more easy, consequently, inserting \cref{eq:157} in \cref{eq:156}, we obtain:
\begin{multline}
- 2\alpha_3\int_{\Sigma_0} \left[R\right]\left\{{g^{\mu\nu}}(\delta {{\Gamma^\lambda}_{\mu\nu}}) - {g^{\mu\nu}}(\delta {{\Gamma^\lambda}_{\mu\lambda}})\right\}\,\sqrt{\lvert\gamma\rvert}\,d^3X\\
- 2\alpha_3\int \left\{{H_\lambda}{\delta {{\Gamma^\lambda}_{\mu\nu}}} - {H_\nu}{\delta {{\Gamma^\lambda}_{\mu\lambda}}}\right\}{g^{\mu\nu}}\,\sqrt{-g}\,d^4X
\end{multline}
\newpage
I shall now introduce the final results for this coefficient:\newline
$\rightarrow\hspace{3mm}(\delta {g_{ij}})_{,n}$:
\begin{align}
\epsilon\left(2{g^{lp}}{g^{ij}}\left[{K_{lp,n}}\right] + 3\left[{A_{n,n}}\right]\right)(\delta {g_{ij}})_{,n}
\label{eq:al3gijn}
\end{align}
$\rightarrow\hspace{3mm}\delta {g_{nn}}$:
\begin{dmath}
\epsilon\left(K + \frac{3}{2}{A_{n}}\right)\left(2{g^{pl}}\left[{K_{lp,n}}\right] + 3\left[{A_{n,n}}\right]\right)(\delta {g_{nn}})
\label{eq:al3gnn}
\end{dmath}
$\rightarrow\hspace{3mm}\delta {g_{ni}}$:
\begin{dmath}
\left\{2\left(2{g^{lp}}{g^{ij}}\left[{K_{lp,n|j}}\right] + 3\left[{A_{n,n|j}}\right]\right) +3{A^{i}}\left(2{g^{pl}}\left[{K_{lp,n}}\right] + 3\left[{A_{n,n}}\right]\right)\right\}(\delta{g_{ni}})
\label{eq:al3gni}
\end{dmath}
With these results at hand, we combine all our pertinent variation coefficients into a single easy to refer to page and proceed to implement them on a special case to obtain their possible physical implications and any conclusions which might lie therein.
\section{All the variation coefficients coalesced}
We use \cref{eq:107,eq:al1ai,eq:al1an,eq:al1gijn,eq:al1gni,eq:al1gnn,eq:al2ai,eq:al2an,eq:al2gijn,eq:al2gni,eq:al2gnn,eq:al3gijn,eq:al3gni,eq:al3gnn}, to obtain this form:\newline
\label{eq:coalesce}
$\rightarrow\hspace{3mm}\delta {A_{n}}$:
\begin{align}
-\left(2\alpha_1 +\alpha_2\right)\left\{4{g^{pl}}\left[{K_{lp,n}}\right] + 6\left[{A_{n,n}}\right]\right\}(\delta {A_n})
\label{eq:allan}
\end{align}
$\rightarrow\hspace{3mm}\delta {A_{i}}$:
\begin{align}
-\left\{\left(3\alpha_1 +5\alpha_2 +4\alpha_4\right)\left[{F^{ni}}\right]\right\}(\delta {A_i})
\label{eq:allai}
\end{align}
$\rightarrow\hspace{3mm}(\delta {g_{ij}})_{,n}$:
\begin{align}
\left\{ \left\{({g^{il}}{g^{jp}})\left(4\alpha_1 +\alpha_2\right) + ({g^{ij}}{g^{lp}})\left(\alpha_2 -4\epsilon\alpha_3\right)\right\}\left[{K_{lp,n}}\right] +\left\{2{g^{ij}}\left(\alpha_1 +\alpha_2\right) - 3\epsilon\alpha_3\right\}\left[{A_{n,n}}\right]\right\}(\delta {g_{ij}})_{,n}
\label{eq:allgijn}
\end{align}
$\rightarrow\hspace{3mm}\delta {g_{nn}}$:
\begin{dmath}
\epsilon\left\{ \left({\alpha_1}(1 + \epsilon) + \frac{\alpha_2}{2}\right)\left[{F^{nl}}_{|l}\right] +\left({A_{l}}{\alpha_1}\right)\left[{F^{nl}}\right] - \left(\frac{3}{4}{A_{l}}{\alpha_2}\right)\left[{F^{ni}}\right] +\left((4{K^{lp}} - {A_n}{g^{lp}}){\alpha_1} + (2K +3{A_n}){g^{lp}}{\alpha_2} -4(K + \frac{3}{2}{A_n}){g^{lp}}{\alpha_3}\right)\left[{K_{lp,n}}\right] +\left((2K + \frac{3}{2}{\epsilon A_n}){\alpha_1} + \frac{3}{2}(2K +3{A_n}){\alpha_2} - (K + \frac{3}{2}{A_n}){\alpha_3}\right)\left[{A_{n,n}}\right]   \right\}(\delta {g_{nn}})
\label{eq:allgnn}
\end{dmath}
$\rightarrow\hspace{3mm}\delta {g_{ni}}$:
\begin{dmath}
- \left\{2({g^{il}}{g^{jp}})(6{\alpha_1 - {\alpha_2} + 4{\alpha_3}})\left[{K_{lp,n|j}}\right]\\ + \frac{1}{2}{g^{ij}}(12{\alpha_1} - {\alpha_2} + 24{\alpha_3})\left[{A_{n,n|j}}\right]\\ + 2{A^{i}}{g^{lp}}(4{\alpha_1} + {\alpha_2} +6{\alpha_3})\left[{K_{lp,n}}\right]\\ +3{A^{i}}(6{\alpha_1} + {\alpha_2} + 6{\alpha_3})\left[{A_{n,n}}\right]\\ - {\epsilon}{{K^{i}}_{l}}(15{\alpha_1} + 2{\alpha_2})\left[{F^{nl}}\right]\\ +3{\epsilon}(2K{\alpha_1} + {A_{n}}{\alpha_2})\left[{F^{ni}}\right]\right\}(\delta {g_{ni}})
\label{eq:allgni}
\end{dmath}
\chapter{Discerning the physics governing the Double Layer}
\label{chap:conclude}
\textbf{About novel surface energy terms:}\newline
We set out on our study of Double layer theory in order to find out their physical implications, whether they manifest the esoteric surface energy on the R.H.S ($S^{nn}$ and $S^{ni}$), which are notably absent in GR. As we see in our final form of the variation coefficients, we find that the coefficients of $\delta g_{nn}$ and $\delta g_{ni}$, which are related to these, since they're part of the same variation, on the L.H.S of our action integral, are not null and void. At this stage, we do not wish to establish precise quantitative factors through which they are related, albeit just to establish that:
\begin{align}
(\cdots){\delta g_{nn}} &= (S^{nn}){\delta g_{nn}}\\
(\cdots){\delta g_{ni}} &= (S^{ni}){\delta g_{ni}}
\end{align}
In GR, these are completely absent. Whereas, in our case, as we see from \eqref{eq:allgnn} and \eqref{eq:allgni}, these are non-vanishing in our premise of Weyl Gravity.\\
$\bullet\,\,{S^{ni}}:$\\
These are promising signs for the physical applicability of Double Layers, because, the term $S^{ni}$ represents the surface energy tensor which (brought to life now, by virtue of these $\delta g_{nn}$ and $\delta g_{ni}$ coefficients) describes how the surface might radiate this energy, from the surface, along the outward normal in turn causing the Double Layer to decay, this can be interpreted one of two ways:
\begin{enumerate}[label = \roman*.]
 \item If we consider our Double Layer to be part of a timelike surface, it can represent a gravitational wave. the surface energy being manifested as a purely gravitational wave, propagating through space.
 \item If we consider our Double Layer to be a part of a spacelike surface, for instance, a typical Schwarzschild Blackhole event-horizon, this surface energy manifest itself through novel particle/matter creation mechanism (we have to investigate the quantum side of things in detail to be able to explain this), propagating through time, like an initial singularity event, for instance the popular "big bang" model.
\end{enumerate}
Of course, all of these contingencies have to investigated further in elaborate mathematical detail, in order to establish anything solid, but the mere presence of a non-zero coefficient for $S^{ni}$ is extremely heartening and can turn out to be a very intriguing case to consider.\newpage
$\bullet\,\,{S^{nn}}:$\\
The $S^{nn}$ component's true nature is largely unknown as of now and we intend to work on it to reveal it's physical properties and any  effects which it might portray.
 \newline
\textbf{About the qualitative relationship between the gauging one-form $A_{n}$ and the extrinsic curvature tensor $K_{lp}$:}\newline
We now talk about the plausible relationship between the gauging one-form and the extrinsic curvature tensor. 
Let us first consider the action principle:
\begin{align}
S_{M} = \int \mathcal{L}_{M}\,\sqrt{-g}\,d^4 x
\end{align}
Further
\begin{align}
\delta S_{M} &= \int \left(\delta\mathcal{L}_{M} + \frac{\mathcal{L}}{2}{g^{\mu\nu}}{\delta g_{\mu\nu}}\right)\,\sqrt{-g}\,d^4 x\\ \nonumber
&= \int\left(\frac{\partial\mathcal{L}_{M}}{\partial{g_{\mu\nu}}}{\delta{g_{\mu\nu}}} + \frac{\mathcal{L}_{M}}{2}{g^{\mu\nu}}{\delta g_{\mu\nu}}\right)\,\sqrt{-g}\,d^4 x\\ \nonumber
&+ \int \frac{\partial\mathcal{L}_{M}}{\partial{{\Gamma^{\lambda}}_{\mu\nu}}}({{\Gamma^{\lambda}}_{\mu\nu}})\,\sqrt{-g}\,d^4 x\\ \nonumber
&= \int \left\{ \left(\frac{\partial\mathcal{L}_{M}}{\partial{g_{\mu\nu}}} + \frac{1}{2}{\mathcal{L}_{M}}{g^{\mu\nu}}\right){\delta{g_{\mu\nu}}} + \frac{\partial\mathcal{L}_{M}}{\partial{{\Gamma^{\lambda}}_{\mu\nu}}}(\delta{{C^{\lambda}}_{\mu\nu}})\right\}\,\sqrt{-g}\,d^4 x\\ \nonumber
&+ \int \frac{\partial\mathcal{L}_{M}}{\partial{{\Gamma^{\lambda}}_{\mu\nu}}}(\delta{{W^{\lambda}}_{\mu\nu}})\,\sqrt{-g}\,d^4 x
\end{align}
We know the variations $(\delta{{C^{\lambda}}_{\mu\nu}})$ and $(\delta{{W^{\lambda}}_{\mu\nu}})$, from \cref{eq:94,eq:95}, ergo we can write (\textcolor{blue}{6.4}) as:
\begin{dmath}
\delta{S_{M}} = \int \left\{ \left(\frac{\partial\mathcal{L}_{M}}{\partial{g_{\mu\nu}}} + \frac{1}{2}{\mathcal{L}_{M}}{g^{\mu\nu}}\right){\delta{g_{\mu\nu}}}\\ + \frac{\partial\mathcal{L}_{M}}{\partial{{\Gamma^{\lambda}}_{\mu\nu}}}\left( (\delta {{C^{\lambda}}_{\mu\nu}}) - {A^{\lambda}}(\delta g_{\mu\nu}) + {A^{\beta}}{g^{\lambda\alpha}}{g_{\mu\nu}}(\delta g_{\alpha\beta})\right) \right\}\,\sqrt{-g}\,d^4 x
+ \int \frac{1}{2}\frac{\partial\mathcal{L}_{M}}{\partial{{\Gamma^{\lambda}}_{\mu\nu}}}\left({g^{\lambda\kappa}}{g_{\mu\nu}}(\delta{A_\kappa}) - {\delta^\lambda_\nu}(\delta{A_\mu}) - {\delta^\lambda_\mu}(\delta{A_\nu})\right)\,\sqrt{-g}\,d^4 x
\label{eq:spl2}
\end{dmath}
As we know, we defined the matter action as \eqref{eq:85}, let's consider the 4-dimensional case of:
\begin{align}
\delta S_{M} \stackrel{def.}{=} -\frac{1}{2} \int {S^{\mu\nu}} (\delta g_{\mu\nu})\,\sqrt{-g}\,d^4X - \int G^\mu (\delta A_\mu)\,\sqrt{-g}\,d^4 x
\label{eq:spl3}
\end{align}
\begin{align}
T^{\mu\nu} = S^{\mu\nu} \delta(n) + \cdots \label{eq:spl4}
\end{align}
\begin{align}
G^{\mu} = Q^{\mu} \delta(n) + \cdots \label{eq:spl5}
\end{align}
These have all been defined before, but let me rewrite them here for clarity:
\begin{align}
\delta S_{M}\big|_{\Sigma_0} = -\frac{1}{2} \int_{\Sigma_0} {S^{\mu\nu}} (\delta g_{\mu\nu})\, \sqrt{\lvert\gamma\rvert}\, {d^3 X} - \int_{\Sigma_0} Q^\mu (\delta A_\mu)\, \sqrt{\lvert\gamma\rvert}\, {d^3 X}\label{eq:spl6}
\end{align}
From this we can infer that the following vanish:\newline
$\delta {A_{n}}$:
\begin{align}
-\left(2\alpha_1 +\alpha_2\right)\left\{4{g^{pl}}\left[{K_{lp,n}}\right] + 6\left[{A_{n,n}}\right]\right\}(\delta {A_n})
\end{align}
$\delta {A_{i}}$:
\begin{align}
-\left\{\left(3\alpha_1 +5\alpha_2 +4\alpha_4\right)\left[{F^{ni}}\right]\right\}(\delta {A_i})
\end{align}
In general we would have: $Q^n \neq 0$, if we let $Q^i=0$ then we would have $\left[F^{ni}\right] = 0$.\\
Now, we can consider the vacuum case wherein, with $Q^n=0$, $Q^i=0$, we are yielded a very interesting relation, wherein:
\begin{align}
-2\left(2\alpha_1 +\alpha_2\right)\left\{2{g^{pl}}\left[{K_{lp,n}}\right] + 3\left[{A_{n,n}}\right]\right\} = 0
\end{align}
The alphas can be non-vanishing, giving us: 
\begin{align}
\frac{2}{3}{g^{pl}}\left[{K_{lp,n}}\right] = -\left[{A_{n,n}}\right]
\end{align}
This relation is a solid link between $\left[{A_{n,n}}\right]$ and $\left[{K_{,n}}\right]$, which is also absent from GR and is a point of stark contrast between Weyl Gravity and GR, a profoundly intriguing highlight into their distinctly differing physical nature.
Although all these inferences are largely qualitative, they provide profound insights into new paths which the Double Layers lead us down. I am hopeful that further research on this subject shall unearth a rigorous mathematical backing to these novel findings. This concludes my thesis. \footnote{As an ending note, I wish to mention a few more references which haven't been mentioned explicitly, but which were of great help in the process of writing this thesis and make for excellent reading: \cite{Berezin2019},\cite{Senovilla2015},\cite{Reina2016}, \cite{alvarez2018}, \cite{Weinberg1974tw}, \cite{eiroa2017}, \cite{salvio2018}, \cite{deBerredoPeixoto2004if}, \cite{Avramidi1986mj}, \cite{Salvio2017qkx}, \cite{doi10106311724264},\cite{Ostrogradsky1850fid}, \cite{CauchyProbWeyl}, \cite{WeylandSpace}}
\label{bib} 
\bibliographystyle{abbrv} 
\bibliography{mybib} 
\end{document}